\documentclass[useAMS,usenatbib,a4paper]{mn2e}
\usepackage{tikz}
\usepackage{amssymb}
\usepackage[normalem]{ulem}

\newcommand{\sph}{{\sc sph}}   
\newcommand{\Msolar}{{\rm M_{\odot}}}   
\newcommand{\Lsolar}{{\rm L_{\odot}}}   
\newcommand{\MJup}{\rm M_{J}}  

\title[Triggered fragmentation in GI discs]{Triggered fragmentation in self-gravitating discs: forming fragments at small radii}
\author[Farzana Meru]{Farzana Meru$^{1,2,3,4}$\thanks{farzana.meru@ast.cam.ac.uk}\\
$^1$Institute of Astronomy, University of Cambridge, Madingley Road, Cambridge, CB3 0HA, United Kingdom\\
$^2$Institut f\"ur Astronomie, ETH Z\"urich, Wolfgang-Pauli-Strasse 27, 8093 Z\"urich, Switzerland\\
$^3$Institut f\"ur Astronomie und Astrophysik, Universit\"at T\"ubingen, Auf der Morgenstelle 10, 72076 T\"ubingen, Germany\\
$^4$School of Physics, University of Exeter, Stocker Road, Exeter, EX4 4QL}
\begin{document}

\maketitle

\label{firstpage}

\begin{abstract}
We carry out three dimensional radiation hydrodynamical simulations of gravitationally unstable discs to explore the movement of mass in a disc following its initial fragmentation.  We find that the radial velocity of the gas in some parts of the disc increases by up to a factor of $\approx 10$ after the disc fragments, compared to before.  While the movement of mass occurs in both the inward and outward directions, the inwards movement can cause the inner spirals of a self-gravitating disc to become sufficiently dense such that they can potentially fragment.  This suggests that the dynamical behaviour of fragmented discs may cause subsequent fragmentation to occur at smaller radii than initially expected, but only \emph{after} an initial fragment has formed in the outer disc.
\end{abstract}


\section{Introduction}
\label{sec:intro}

Self-gravitating discs are known to be important on a wide range of astrophysical scales.  If they fragment, a number of objects may potentially form out of discs of different sizes including stars \citep[e.g.][]{Nayakshin_GI_AGN}, brown dwarfs \citep[e.g.][]{Stamatellos_BD_formation} and planets \citep[e.g.][]{Boss_GI,Mayer_GI_Sci}.

The stability of a self-gravitating disc can be described by the stability parameter \citep{Safronov_GI,Toomre_stability1964},

\begin{equation}
Q = \frac{c_s \kappa}{\pi \Sigma G},
\end{equation}
where $c_{\rm s}$ is the sound speed in the disc, $\kappa_{\rm ep}$ is the epicyclic frequency, which for Keplerian discs is approximately equal to the angular frequency, $\Omega$, $\Sigma$ is the surface mass density and $G$ is the gravitational constant.  For an infinitesimally thin disc to fragment, the stability parameter must be less than a critical value, $Q_{\rm crit} \approx 1$ (\citealp{Safronov_GI,Toomre_stability1964}; see \citealp{Helled_PPVI} for a recent review).

Investigations into gravitational instability typically consider discs up until the fragmentation stage.  Those simulations that do consider the subsequent evolution of such discs focus on the properties of the fragments that form and the longer term state of the system \citep[e.g.][]{Mayer_GI_Sci,Rice_frag_evol,Stamatellos_BD,Stamatellos2011_obs_discs,Zhu_GI_migration,Vorobyov_gaps}.  Following fragmentation there are a number of processes that can take place.  These may affect the fragment (e.g. collapse, accretion, metal enrichment, tidal disruption, migration \& gap-opening; \citealp[][]{Galvagni_clump_collapse,Boley_GI_metal_enrichment,Boley_tidal_downsizing,Nayakshin_tidal_downsizing,GI_migration,Malik_Meru_a}), the star (e.g. FU Orionis outbursts; \citealp{Lodato_Clarke_FUOri,Vorobyov_Basu_GI_FUOri,Vorobyov_Basu_GI_FUOri2}) and the disc (e.g. loss of spiral structure, gap-opening; \citealp[][]{Stamatellos_BD,Meru_PhD,Malik_Meru_a}).  However, little is understood specifically about the movement of the mass in the disc following the formation of the first fragment and how this affects the possibility of subsequent fragmentation.  In the context of planet formation \cite{Armitage_Hansen_trigger} performed Smoothed Particle Hydrodynamics simulations of self-gravitating discs and inserted a planet (using a smoothed point mass) to explore the subsequent effect on the disc.  They found that the presence of a planet triggers the formation of fragments both exterior and interior to it.  These simulations, however, considered an isothermal equation of state, which favours fragmentation.

In this paper we use 3D SPH simulations with radiative transfer to self-consistently model the fragmentation of a young self-gravitating disc and the subsequent movement of gas in the disc.  We show that the formation of the first fragment causes the radial velocity of the disc mass to increase (by up to a factor of $\approx 10$) in both the inwards and outwards directions.  The inwards movement, which occurs \emph{because} of the presence of the first fragment, can cause the inner regions of a disc that are either stable or marginally stable (so that $Q \approx 1$) to become sufficiently dense and unstable such that they may potentially fragment.  

In Section~\ref{sec:numerics} we describe our numerical method and present the details of the simulations in Section~\ref{sec:sim}.  We present our results in Section~\ref{sec:results} and discuss and make conclusions in Sections~\ref{sec:disc} and~\ref{sec:conc}, respectively.

\section{Numerical method}
\label{sec:numerics} 

The Smoothed Particle Hydrodynamics (\sph) code used to carry out the simulations presented here is almost exactly the same as the code used by \cite{Meru_Bate_opacity}, originally developed by \cite{Benz1990} and further developed by \cite*{Bate_Bonnell_Price_sink_ptcls}, \cite*{WH_Bate_Monaghan2005}, \cite{WH_Bate_science} and \cite{Price_Bate_MHD_h}.  We use the flux-limited diffusion approximation to model the radiation transport with two temperatures (that of the gas and the radiation field).  To model the shocks in the discs, we use an artificial viscosity \citep{Monaghan_Gingold_art_vis} with \sph~parameters fixed at $(\alpha_{\rm SPH}, \beta_{\rm SPH}) = (0.1, 0.2)$.  We use grey Rosseland mean opacities based on the values of \cite*{Pollack_etal_dust_opacitites} for interstellar molecular dust grains and \cite{Alexander_gas_opacities} for the gas contributions at high temperatures (see \citealp{WH_Bate_science} for details on the opacity tables used).  We carry out simulations with opacity values that are lower than interstellar values and scale the values in the tables by the required factor as done so by \cite{Meru_Bate_opacity}.

The equation of state used in these simulations assumes that the gas is composed of hydrogen (70\%) and helium (28\%) and includes the rotational and vibrational modes of molecular hydrogen, dissociation of molecular hydrogen and ionisation of both hydrogen and helium.  We use a ratio of specific heats, $\gamma = 5/3$.  Our models  assume that the presence of dust only affects the value of the opacity in the disc.

We model the effects of stellar irradiation using a two-layer approach, as done by \cite{Meru_Bate_opacity}.  The midplane region is modelled using the flux-limited diffusion approximation and the boundary layer of particles maintains a fixed temperature profile (assumed to be set by the radiation field of the star) such that any energy that is passed to these boundary particles is effectively radiated away.  The vertical location of the boundary between the optically thick part of the disc and the optically thin atmosphere is determined slightly differently to \cite{Meru_Bate_opacity}: their boundary location was determined at the start of the simulation and then kept fixed throughout while our method re-evaluates the boundary location at least 33 times per outer rotation period (ORP).  The boundary height is the maximum out of: the height above the midplane where the optical depth, $\tau = 1$, or the height above the midplane where 10\% of the particles at any radius are in the boundary.  Therefore, the boundary is at the vertical position where the optical depth, $\tau \lesssim 1$, and our choice ensures that the ``bulk'' of the disc (i.e. consisting of particles that lie closer to the disc midplane) is simulated using radiative transfer, rather than simplified energetic calculations.  As described by \cite{Meru_Bate_opacity} the particles that make up the boundary do change over time since particles may move across the boundary from the optically thick to optically thin region, or vice versa.

To decrease the computational expense, we turn any fragments into sink particles \citep*{Bate_Bonnell_Price_sink_ptcls} when they reach a critical density of $\rho_{\rm crit} = 7 - 9 \times 10^{-6} \rm g/cm^3$, i.e. denser than that required for second core collapse, allowing us to follow the evolution of the disc further.  In the simulations presented here, the first fragments form at $\approx 33 - 55$~AU and have masses $\approx 1 \MJup$.  This gives a Hill radius (defining the region from which a core typically accretes from) of $\approx 2 - 3$AU.  We therefore model the fragments using an accretion radius of 1~AU i.e. we assume that any gas within 1~AU of the sink particle can be accreted onto it.

\section{Simulations}
\label{sec:sim}

\begin{figure}
  \centering
\includegraphics[width=1.1\columnwidth]{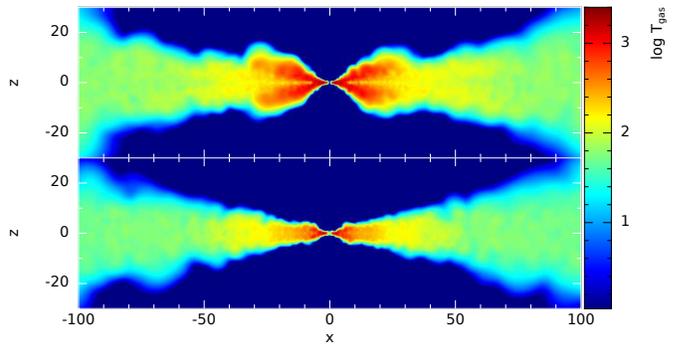}
 \caption{Logarithm of the gas temperature (in K) rendered in cross-sectional views of the simulations before any fragments form.  Top panel: Simulation 1 at $t = 1.29$~ORPs.  Bottom panel: Simulation 2 at $t = 1.92$~ORPs.  Axis units are in au.  The surface temperatures of the discs are $\approx 10 - 50$~K, and the outer disc is close to being vertically isothermal.}
  \label{fig:T}
\end{figure}

We carry out simulations of two discs with $\Sigma \propto R^{-3/2}$ and radial extent $1 < R < 100$~AU.  Since the aim of this paper is to investigate the radial movement of the disc mass following the initial fragmentation, we set up the discs such that they do fragment.  Discs are more likely to fragment at large radii \citep{Rafikov_SI,Clarke2009_analytical,Boley_CA_and_GI}, if the disc mass is high \citep{Rafikov_unrealistic_conditions}, or if the metallicity is low \citep{Cai_etal_RT,Meru_Bate_opacity,Rogers_Wadsley_kappa}.  We assume the initial and boundary temperature profile to be $T \propto R^{-1/2}$ and set the absolute value of the temperature from the luminosity, $L$, of the central star (using $L = 4 \pi R^2 \sigma T^4$, where $\sigma$ is the Stefan-Boltzmann constant).  The luminosity is determined using the stellar evolution models of \cite{Seiss_models} for a low metallicity $Z=0.01$ at an age of 1~Myr (i.e. at a sufficiently early age when a disc would have been present around the central star).  The resulting surface temperature of the discs modelled are $\approx 10 - 50$~K (see Figure~\ref{fig:T}), in agreement with previous similar fragmentation simulations: \cite{Boley_CA_and_GI} sets the disc's irradiation temperature to 30K, \cite{Stamatellos_BD} have one that varies with radius and is O(10)~K, and \cite{Cha_Nayakshin} set their ambient gas temperature to 10K.  It is important to note that while the luminosity may be important for disc fragmentation (since it affects the disc energetics), this investigation is concerned with the dynamics of the disc rather than its thermodynamics. 

\cite{Meru_Bate_resolution,Meru_Bate_convergence} have called into question the detailed thermodynamical conditions under which fragmentation can occur.  However \cite{Lodato_Clarke_resolution} suggest that the results concerning the dynamics of such discs are more likely to be robust since the resolution that is required to accurately model such large-scale processes are not as stringent as that required to resolve fragmentation.  Since we are more concerned with the dynamics of the discs, i.e. how the mass in the disc moves in response to the presence of a fragment and the \emph{potential} for further fragmentation rather than the actual process of fragmentation itself, we simulate these discs using 250,000 particles.

We firstly simulate a $1.1 \Msolar$ disc surrounding a $1.5 \Msolar$ star with luminosity $L = 4.3 \Lsolar$.  We model this disc with opacity values that are 0.9 times the interstellar Rosseland mean values.  We also present the results of a second simulation with a disc mass of $0.9 \Msolar$ around a $1.0 \Msolar$ star with $L = 2.4 \Lsolar$.  To ensure that the disc does initially fragment, we simulate it with a low opacity (0.5 times the interstellar Rosseland mean values).  The setup is such that the minimum value of the stability parameter (at the outer disc edge) is $Q_{\rm min} = 0.75$ and 0.7 in Simulations 1 and 2, respectively.  However, the discs do heat up due to work done and viscous heating and hence the stability parameter does increase from the initial values.

\section{Results}
\label{sec:results}

\begin{figure*}
\centering \includegraphics[width=0.75\columnwidth,angle=-90,origin=br]{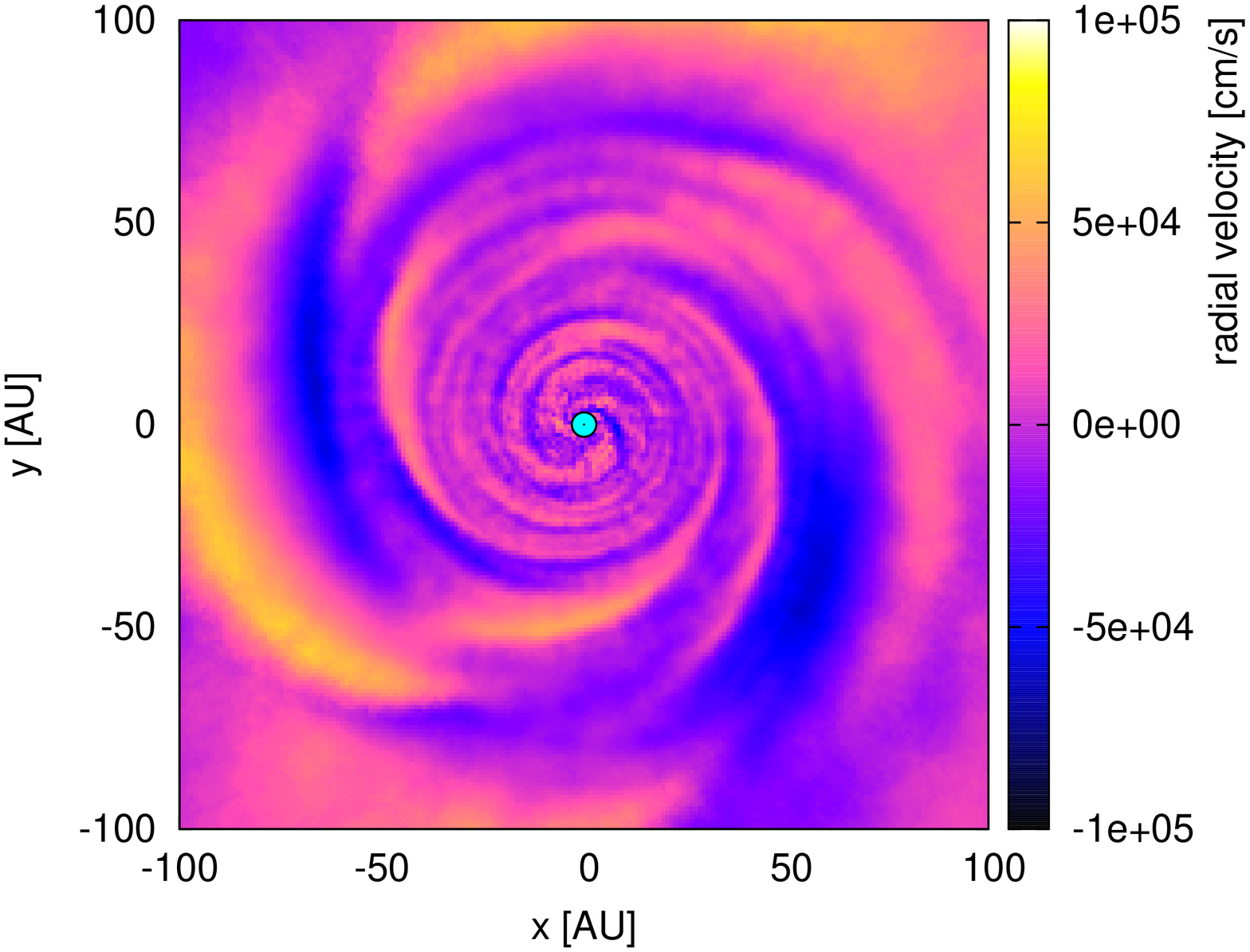}
\includegraphics[width=1.0\columnwidth]{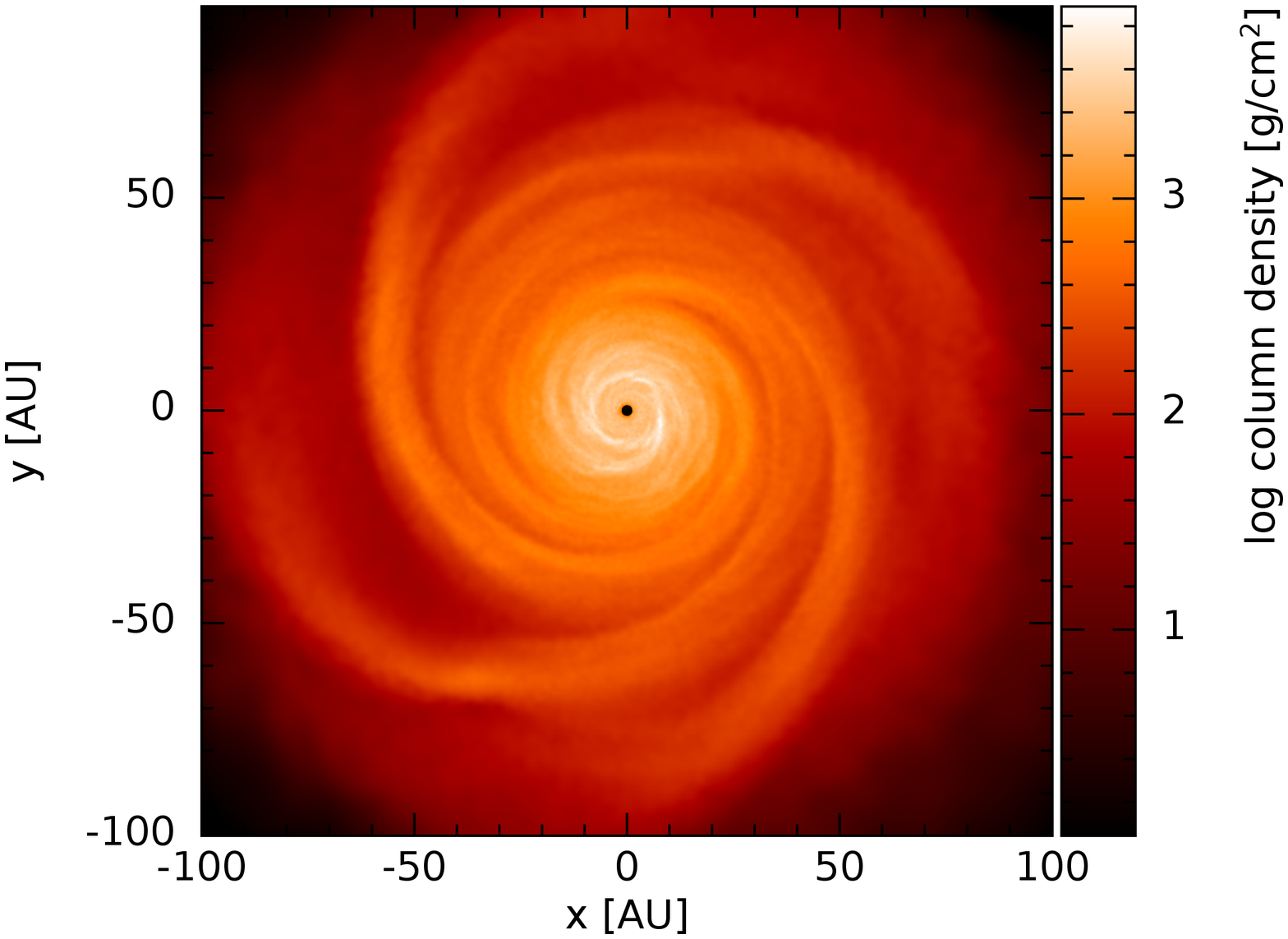}
 \caption{Radial velocity (left panel) and surface mass density (right panel) rendered images of Simulation 1 at a time ($t=1.07$~ORPs), shortly before the formation of the first fragment (at $t=1.30$~ORPs).  At this time the gravitational instability has developed and the radial velocity is neither at one extreme nor the other.  Negative and positive values of the velocity indicate inwards and outwards movement, respectively.  The central star is given by blue and black dots in the left and right panels, respectively.}
\label{fig:sim1_pre_fragmentation}
\end{figure*}

\begin{figure*}
\centering \includegraphics[width=0.75\columnwidth,angle=-90,origin=br]{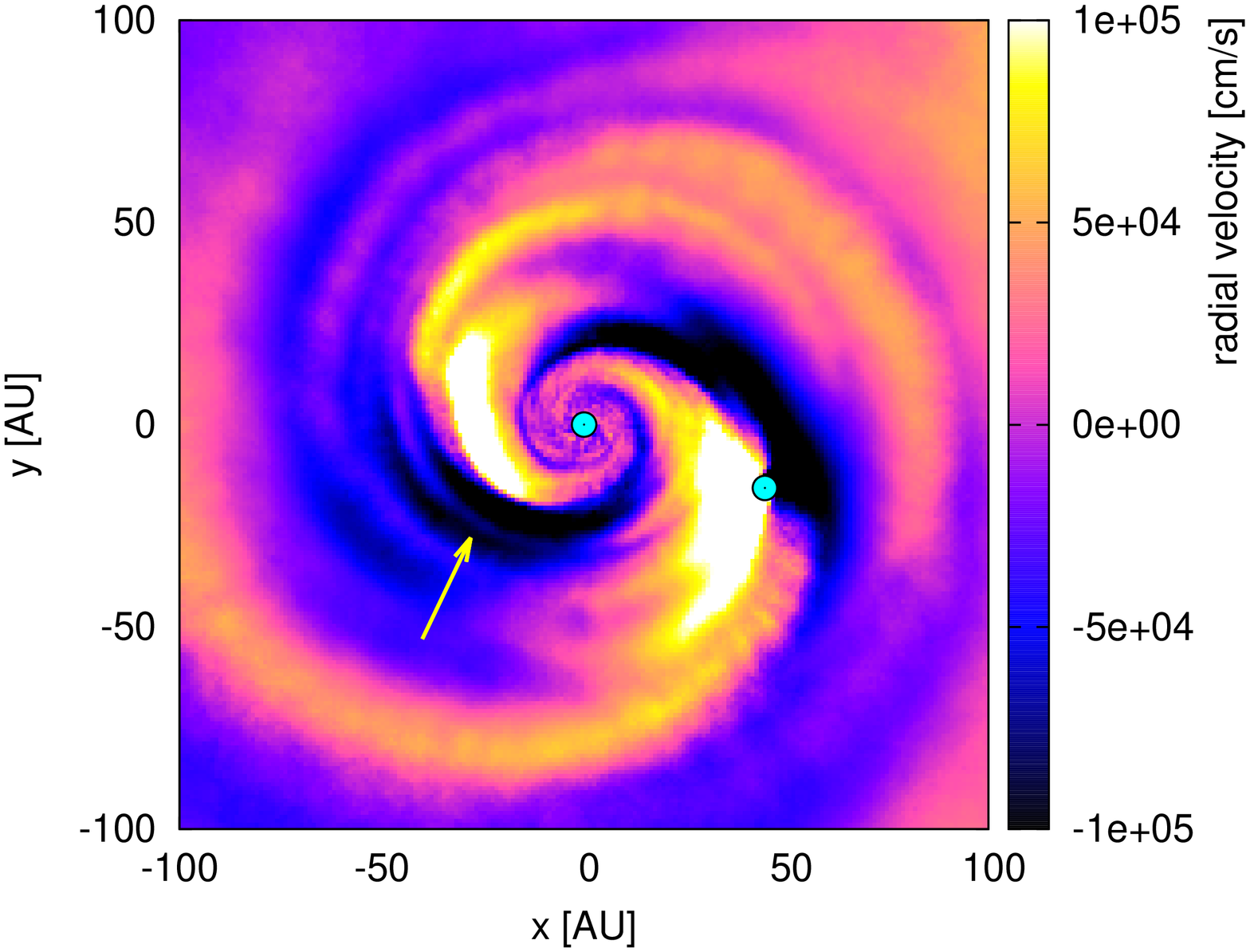}
\includegraphics[width=1.0\columnwidth]{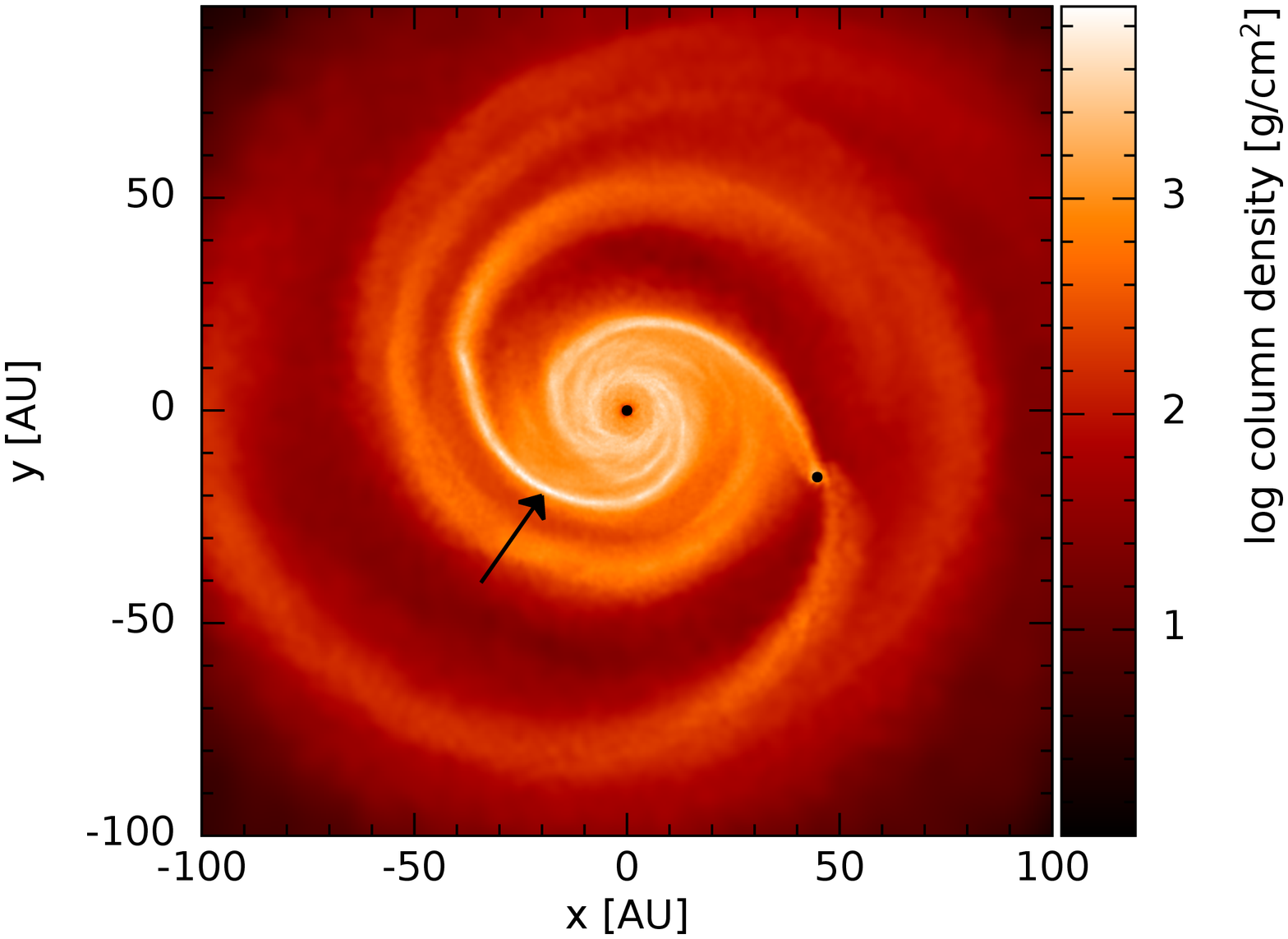}
 \caption{Radial velocity (left panel) and surface mass density (right panel) rendered images of Simulation 1 at a time, $t = 1.75$ ORPs, shortly before the second fragment forms.  After the formation of the first fragment material moves radially inwards (marked by an arrow in the left panel), causing the inner spiral (indicated by an arrow in the right panel) to become sufficiently dense that it can fragment (Figure~\ref{fig:sim1_2fragments}).  The central star and fragment are given by blue and black dots in the left and right panels, respectively.}
\label{fig:sim1_1fragment}
\end{figure*}

\begin{figure}
\centering
\includegraphics[width=1.0\columnwidth]{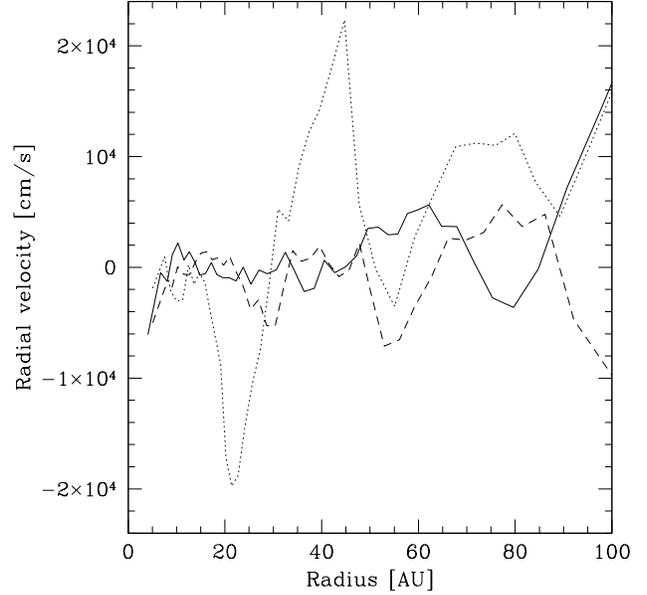}
\caption{Azimuthally averaged radial velocity profile in Simulation 1 shortly before the formation of the first ($t = 1.19$ ORPs; solid line) and second ($t = 1.75$ ORPs; dotted line) fragments.  The radial velocities increase by up to an order of magnitude after the first fragment forms compared to before.  The dashed line shows the radial velocity of the disc at $t = 1.75$ ORPs but in the disc where the formation of the first fragment has been artificially suppressed (Section~\ref{sec:kill}).  Without the presence of the first fragment the movement of material in the disc is no longer enhanced.}
\label{fig:vR}
\end{figure}

\begin{figure*}
\centering \includegraphics[width=0.75\columnwidth,angle=-90,origin=br]{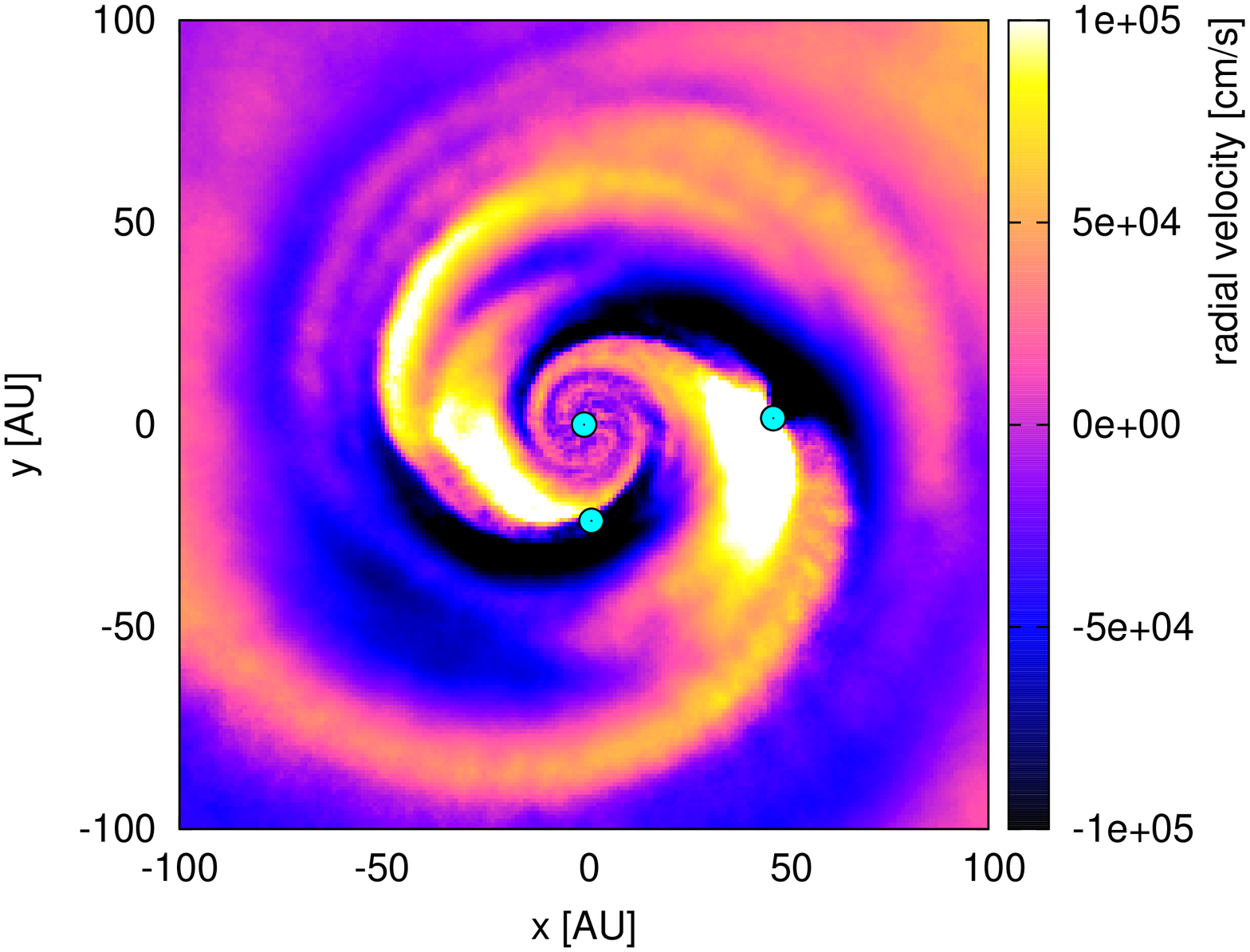}
\includegraphics[width=1.0\columnwidth]{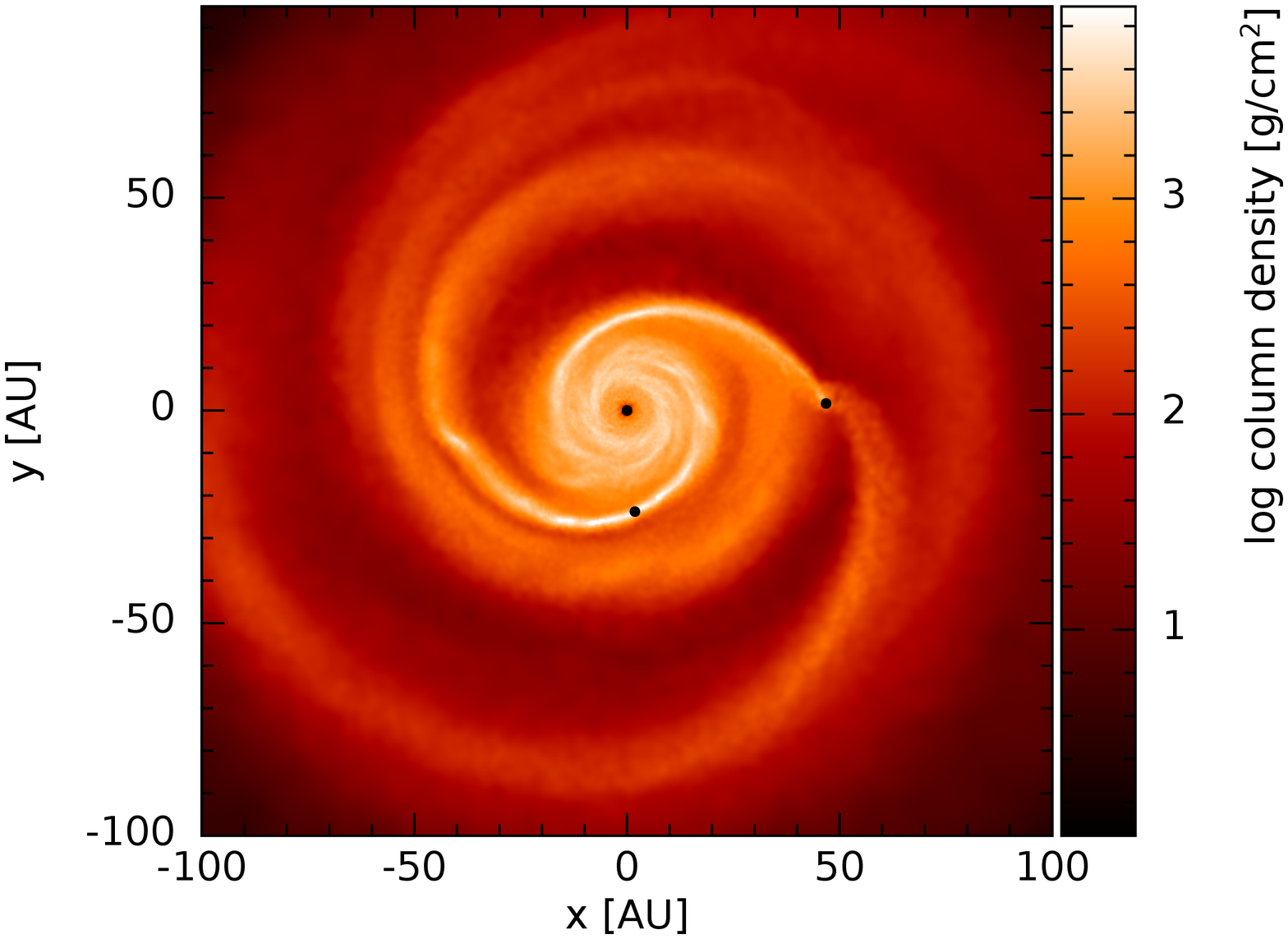}
 \caption{Radial velocity (left panel) and surface mass density (right panel) rendered image of Simulation 1 at the time when the second fragment forms ($t=1.76$~ORPs).  This fragment forms out of a spiral arm that becomes dense due to the movement of mass radially inwards following the formation of the first fragment (Figure~\ref{fig:sim1_1fragment}).  The central star and fragments are given by blue and black dots in the left and right panels, respectively.  Following the formation of the second fragment the disc becomes much more dynamic.  The presence of multiple fragments causes several regions to have radial velocities that are higher than they were prior to fragmentation (c.f. Figure~\ref{fig:sim1_pre_fragmentation} left panel and Figure~\ref{fig:vR}).}
 \label{fig:sim1_2fragments}
\end{figure*}

Figure~\ref{fig:sim1_pre_fragmentation} shows the radial velocity and surface mass density rendered images of the disc in Simulation 1 prior to the formation of any fragments when the gravitational instability has developed.  The radial velocity plots are determined by projecting all the SPH particles onto a single x-y plane.  The left panel shows that the movement of mass in the disc is both inwards and outwards but that the radial velocity of the gas is not extremely large in either direction.  

Figure~\ref{fig:sim1_1fragment} shows the radial velocity and surface mass density rendered images of the disc a short while after the first fragment forms (the fragment forms at $\approx 54.5$~AU at a time $t \approx 1.30$~ORPs).  This figure shows that mass is moving both outwards and inwards in the vicinity of the first fragment (the inwards motion is represented by the black region in the left panel).  The magnitude of the radial velocities are up to a factor of $\approx 10$ larger than before fragmentation (compare solid and dotted lines in Figure~\ref{fig:vR}) and hence the disc is much more dynamic with more extreme mass movement.  This movement into the inner regions causes mass to accumulate onto an inner spiral arm (indicated by an arrow in the right panel of Figure~\ref{fig:sim1_1fragment}).  This spiral arm then fragments a short while later (at $\approx 24$~AU at a time $t=1.76$~ORPs; Figure~\ref{fig:sim1_2fragments}).

\begin{figure*}
\centering \includegraphics[width=0.9\columnwidth]{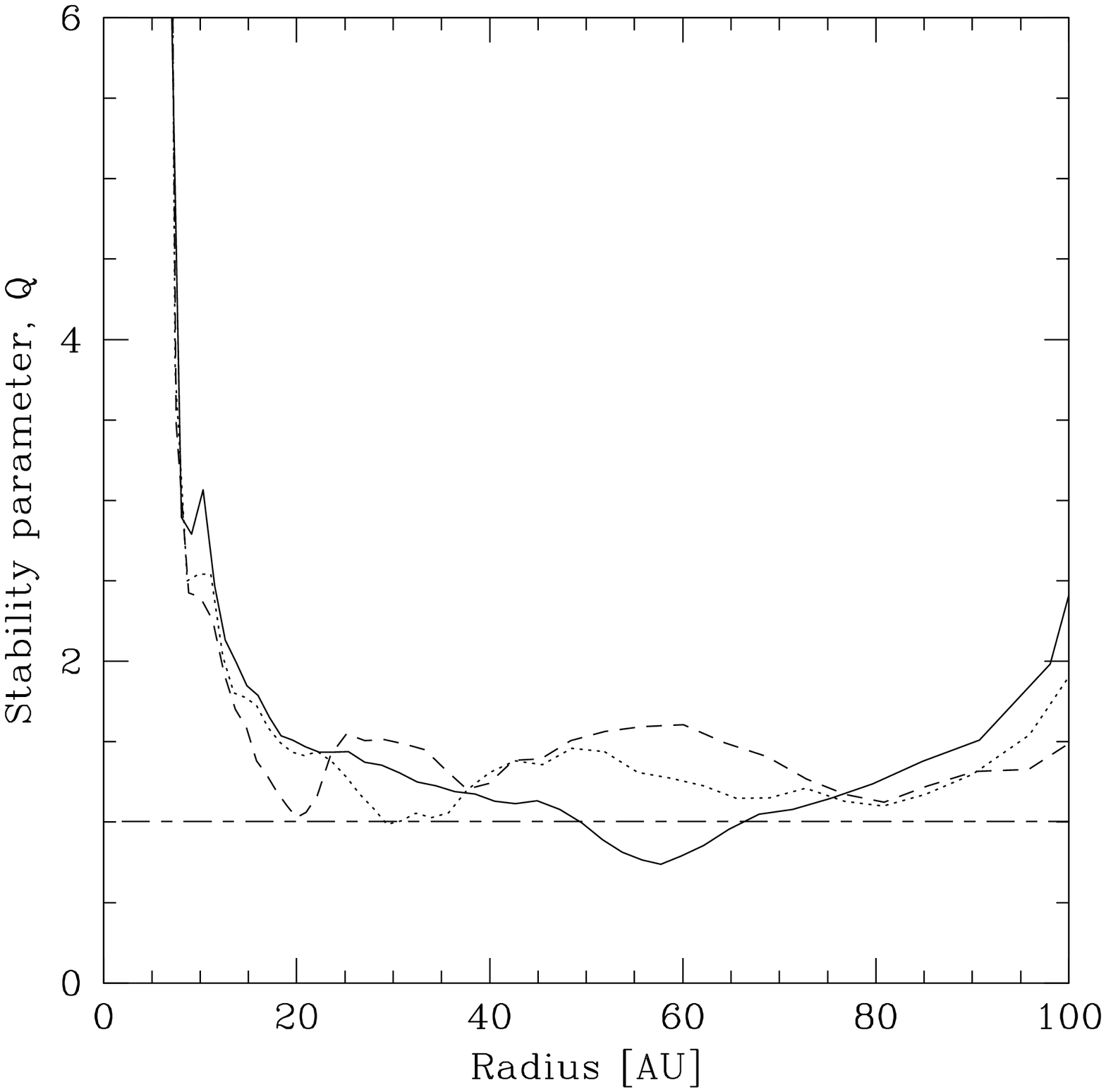}\\
\includegraphics[width=0.73\columnwidth,angle=-90,origin=br]{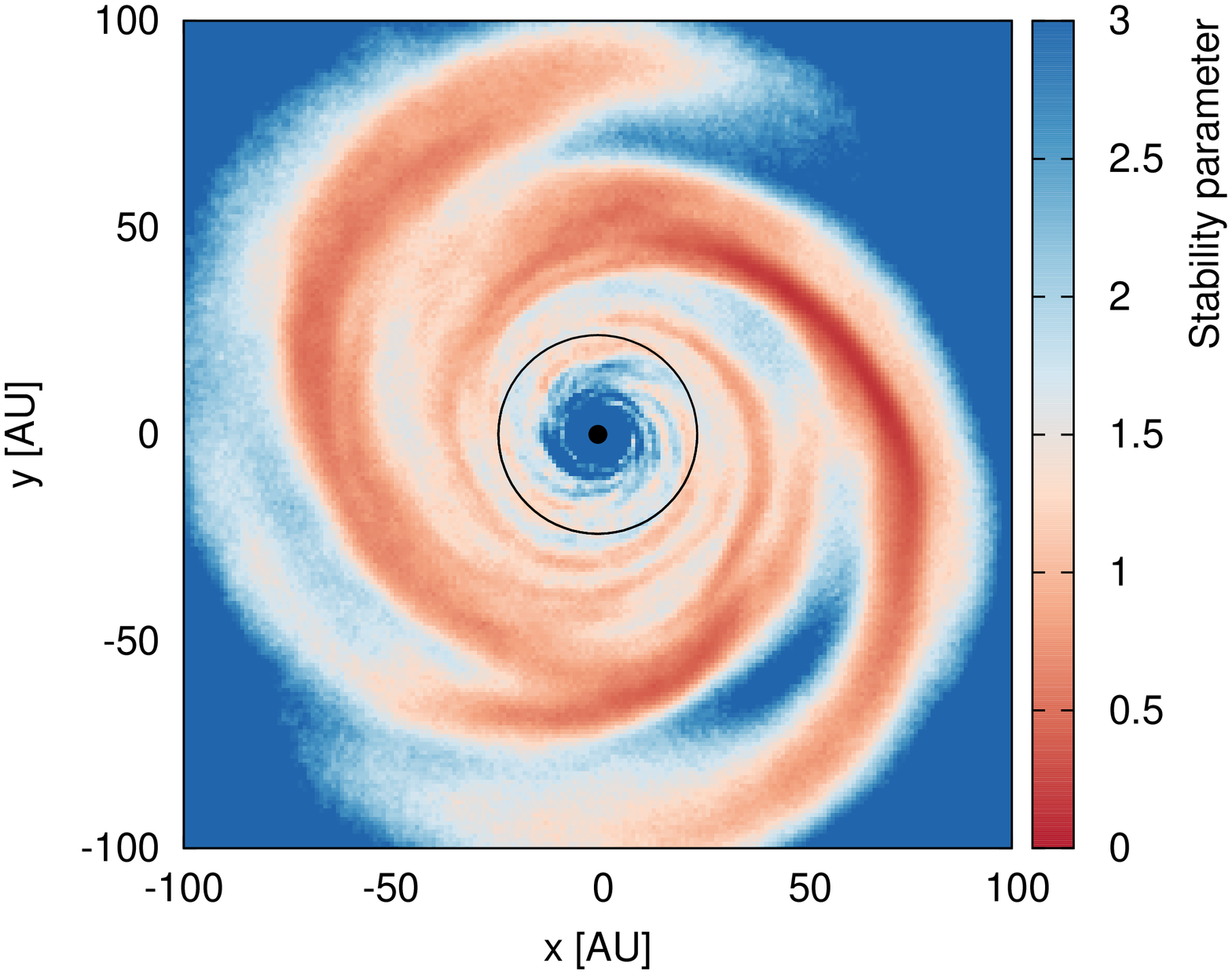}
\includegraphics[width=0.73\columnwidth,angle=-90,origin=br]{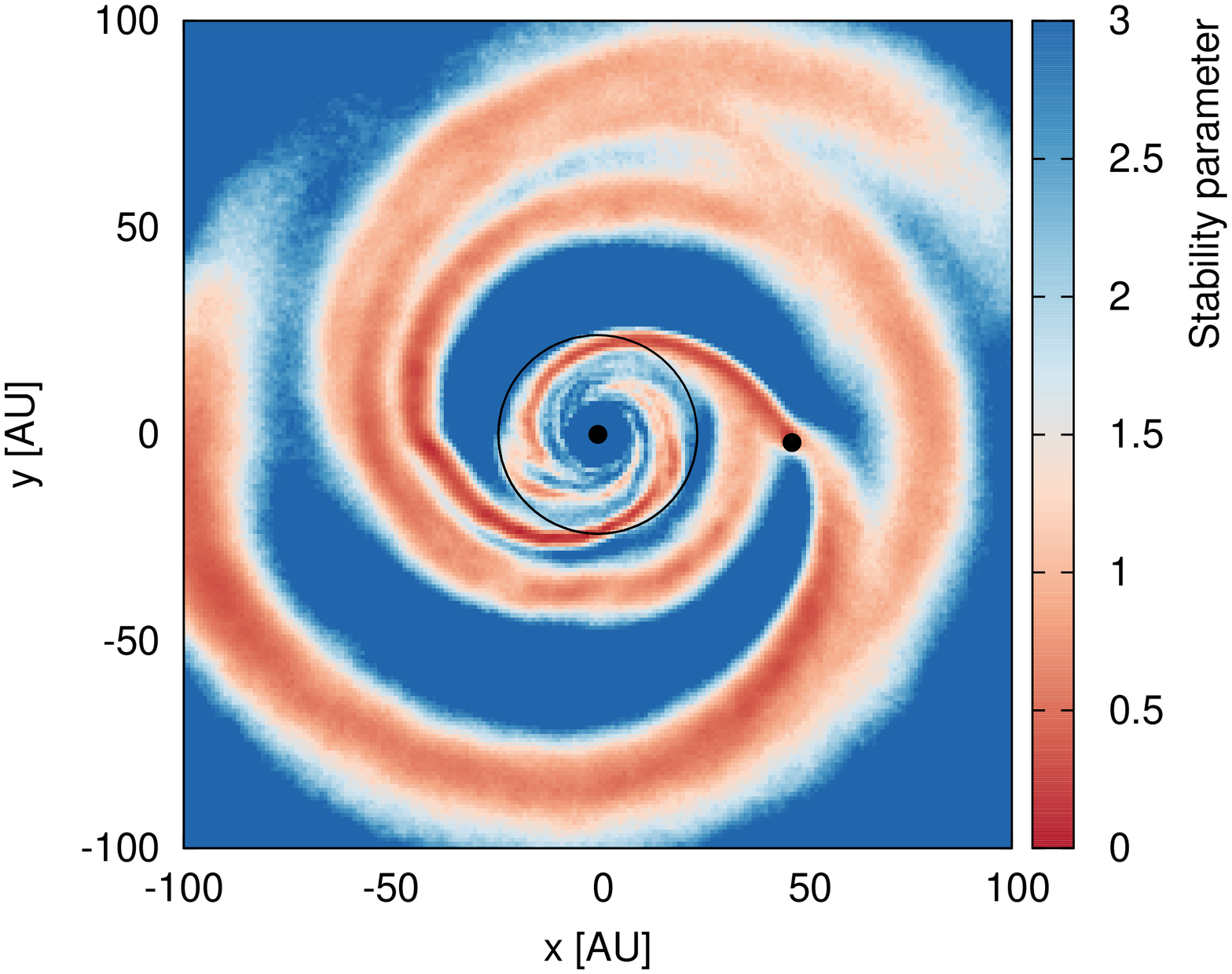}
 \caption{Top panel: Azimuthally averaged stability profiles of the disc in Simulation 1 shortly before the first (solid line; t = 1.19 ORPs) fragment forms and at two instances before the second fragment forms (at t = 1.62 and 1.76 ORPs; dotted and short dashed lines, respectively).  Before any fragments form, part of the outer disc (beyond $\approx 50$AU) is unstable and can fragment while the inner disc at $\approx 24$ au (where the second fragment later forms) is stable ($Q > 1$).  The mass movement into the inner parts of the disc following the formation of the first fragment causes the inner spiral to be pushed into a state of instability such that it fragments (at $\approx 24$~AU).  Note that the stability profile does not show that the disc is unstable at this location.  However such azimuthally averaged plots wash out the local properties.  Bottom panels: Local stability parameter values before the first and second fragments form (at $t \approx  1.19$ and 1.76 ORPs, respectively).  The black circles indicate the radius at 24 au.  Before the first fragment forms the value of the stability parameter is $\gtrsim 1$, and hence stable, inside 50AU (bottom left).  Before the second fragment forms regions of the annulus at $\approx 24$ au are definitely unstable with $Q < 1$ (bottom right).  The second fragmentation is therefore unlikely to have occurred without the aid of the mass movement.  The central star and fragment are given by black dots.}
 \label{fig:sim1_Q}
\end{figure*}

Figure~\ref{fig:sim1_Q} shows the azimuthally averaged stability profile for the disc just before the formation of the first and second fragments, as well as a 2D map of the stability parameter in the disc before the first and second fragments form.  The 2D stability values are calculated using the mass and internal energy of the SPH particles that are closest to the x-y position plotted, i.e. by projecting the SPH particles onto a single x-y plane.  This is sufficient as firstly, the temperature of most of the particles lies within a factor of a few of the midplane temperature, especially in the outer disc which is close to being vertically isothermal (see Figure~\ref{fig:T}).  Secondly, the stability parameter only depends on the square root of the temperature.  Prior to the first fragment forming, regions of the outer disc are gravitationally unstable (the stability parameter, $Q < 1$) but at $R \lesssim 50$~AU, $Q \gtrsim 1$, and thus is not so susceptible to fragmentation.  However, following the formation of the first fragment, the inner spiral becomes sufficiently dense such that its stability parameter decreases, allowing it to be pushed (from a state of marginal stability where $Q > 1$) into a fragmentation state (at $\approx 24$~AU).

\subsection{Would the inner disc have fragmented anyway?}
\label{sec:kill}

\begin{figure}
\centering
\includegraphics[width=1.0\columnwidth]{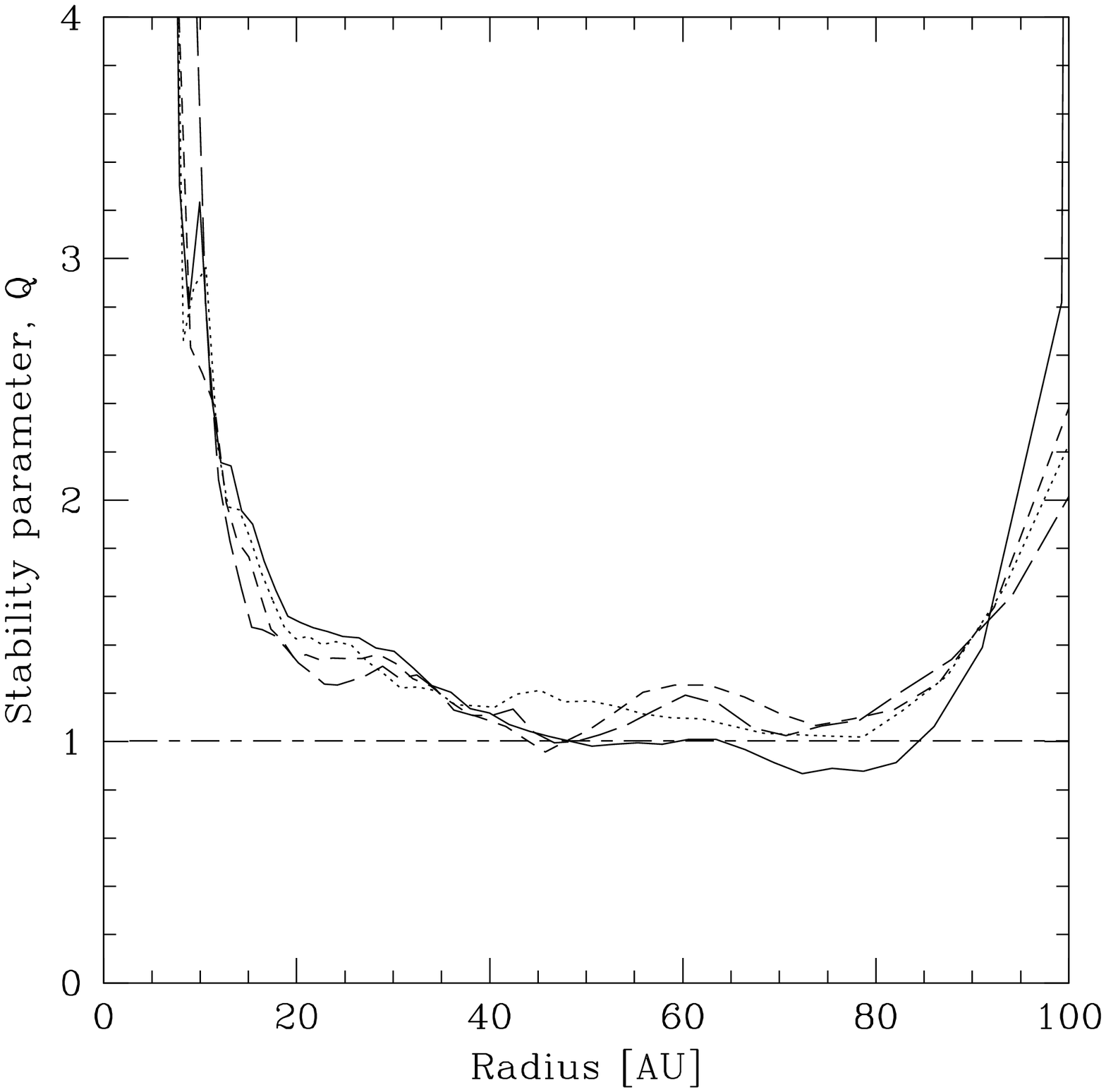}
 \caption{Azimuthally averaged stability profiles of the disc in Simulation 1 that is modified to remove an annulus of particles at various times: just before any particles are removed ($t = 1.04$ ORPs; solid line), at the time when the first fragment formed in the original simulation ($t = 1.30$~ORPs; dotted line), at the time when the second fragment formed in the original simulation ($t = 1.76$~ORPs; short dashed line) and much later in the simulation ($t = 3.0$~ORPs; long dashed line).  The stability parameter does not change significantly in the inner disc at $\approx 24$ au.  This suggests that the fragmentation in the outer disc may be the cause of the inner disc fragmenting in the original simulation.}
 \label{fig:Q_kill}
\end{figure}

\begin{figure}
\centering
\includegraphics[width=0.8\columnwidth,angle=-90]{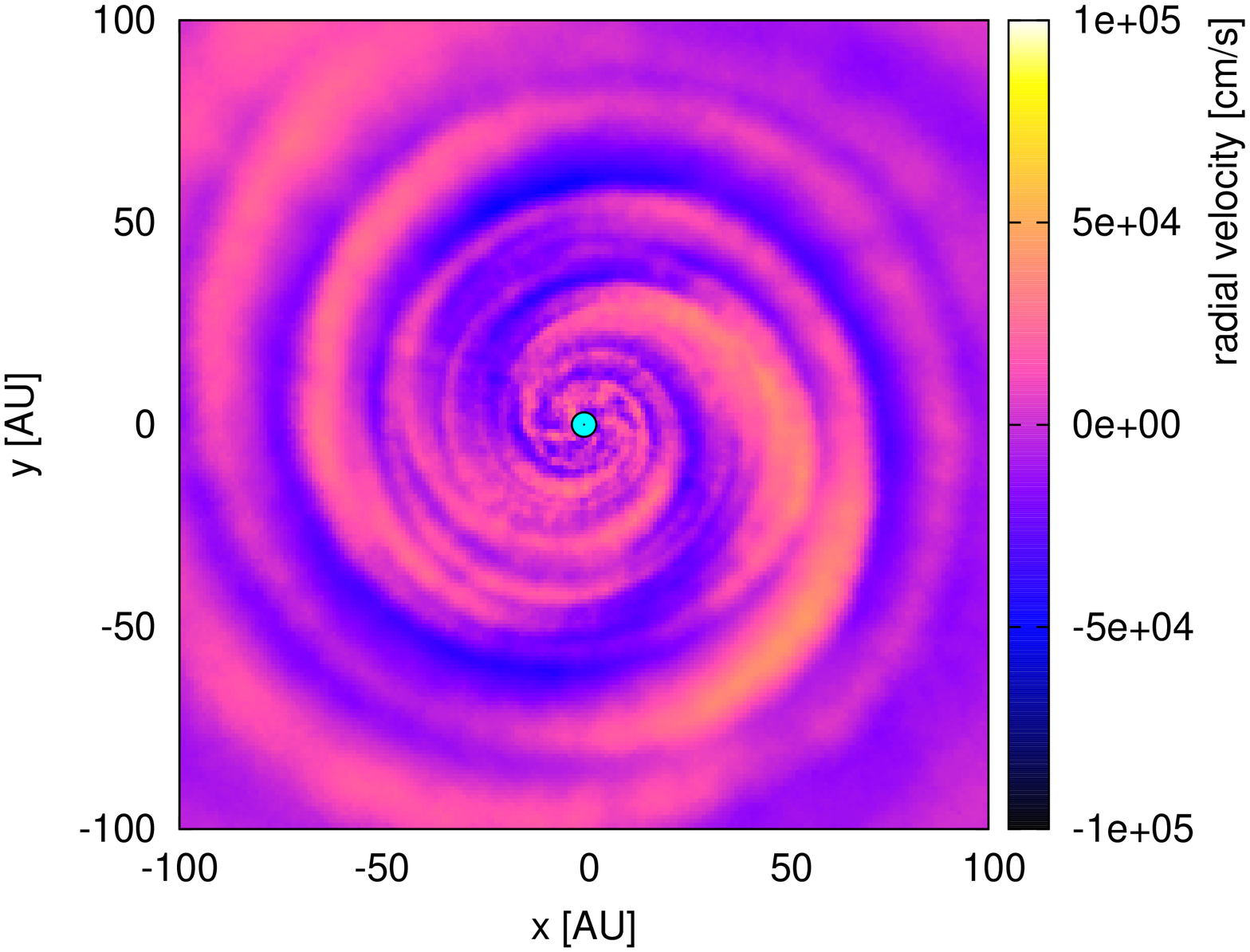}
 \caption{Radial velocity rendered image of the disc in Simulation 1 that is modified to remove an annulus of particles.  The image is at the end of the simulation when no fragments have formed.  The disc is quiescent in the inner regions, similar to the original simulation before fragmentation occurs (Figure~\ref{fig:sim1_pre_fragmentation}, left panel).  The disc is also much more quiescent in the inner disc compared to the original simulation when fragmentation occurs (Figure~\ref{fig:sim1_1fragment}, left panel).}
 \label{fig:vR_end_kill}
\end{figure}

In the above section we suggest that the formation of the second fragment was a consequence of the presence of the first fragment and the resulting movement of disc gas.  However, to show that the second fragment would not have formed regardless of the presence of the first fragment, we restart the simulation at an earlier time, $t = 1.04$~ORPs (shortly before the increase in surface mass density begins at the radius where the first fragment forms), but remove all the particles that are within a 1~au annulus of where the first fragment forms, i.e. the particles within $53.5 < R < 55.5$ au (totalling $0.019\Msolar$ which is 1.8\% of the SPH particles at that time).  This artificially stops the first fragment forming and allows us to see if the second fragment would have formed anyway.  We run the simulation until 8.9 ORPs (i.e. 5 times longer than it took until the second fragment formed in the original simulation) and find that the inner fragment does not form at all.  In addition, the radial velocity rendered images show a quiescent disc, i.e. no mass movement takes place, similar to what is seen before any fragmentation occurs in the original simulation (Figure~\ref{fig:sim1_pre_fragmentation}).

To ensure that this artificial removal of particles from the annulus does not cause the gas from the inner disc to replenish this artificial gap (and thus artificially decrease the surface mass density of gas near to where the inner fragment forms) we observe the stability parameter plot of the inner disc.  Figure~\ref{fig:Q_kill} shows the stability profile of this disc at various times (before the particles are removed and afterwards) and shows that the stability profile at $\approx 24 $ au marginally decreases over time.  If the artificial removal of particles caused material to move away from the disc region at $\approx 24$ au the surface density would decrease and the stability parameter would increase.  The fact that this is not happening shows that the lack of fragmentation cannot be due to this.  This provides further confidence that the reason for the second fragment forming is due to the presence of the first fragment.  Furthermore, Figure~\ref{fig:vR_end_kill} shows the radial velocity rendered plot of the image at the end of the simulation (at $t = 8.9$~ORPs) and shows that the mass movement in the disc is as quiescent in the inner regions as it was before any fragmentation occurred (Figure~\ref{fig:sim1_pre_fragmentation}, left panel).

\subsection{Simulation 2}

\begin{figure*}
\centering 
\vspace{-0.4cm}
\includegraphics[width=0.67\columnwidth,angle=-90,origin=br]{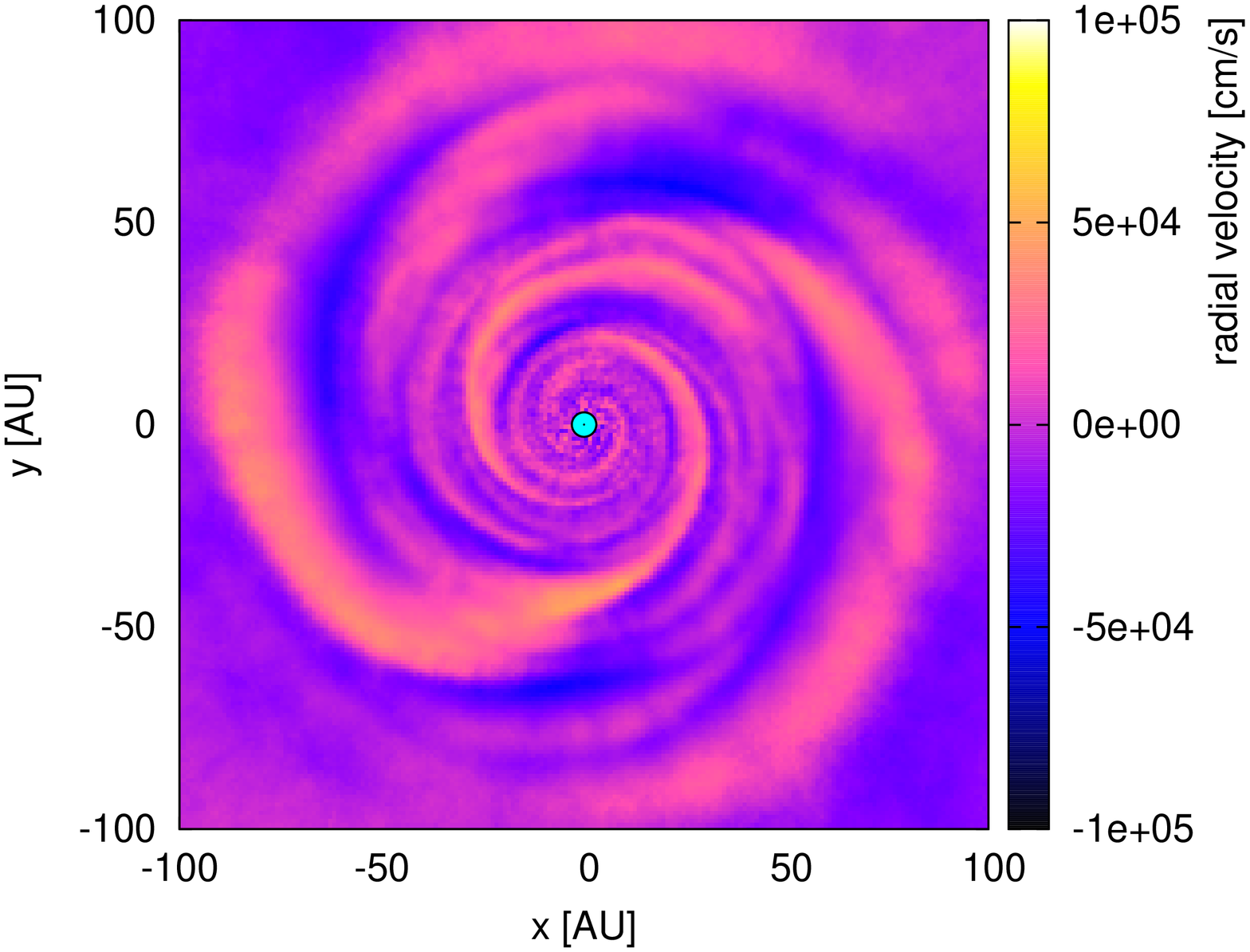}
\includegraphics[width=0.9\columnwidth]{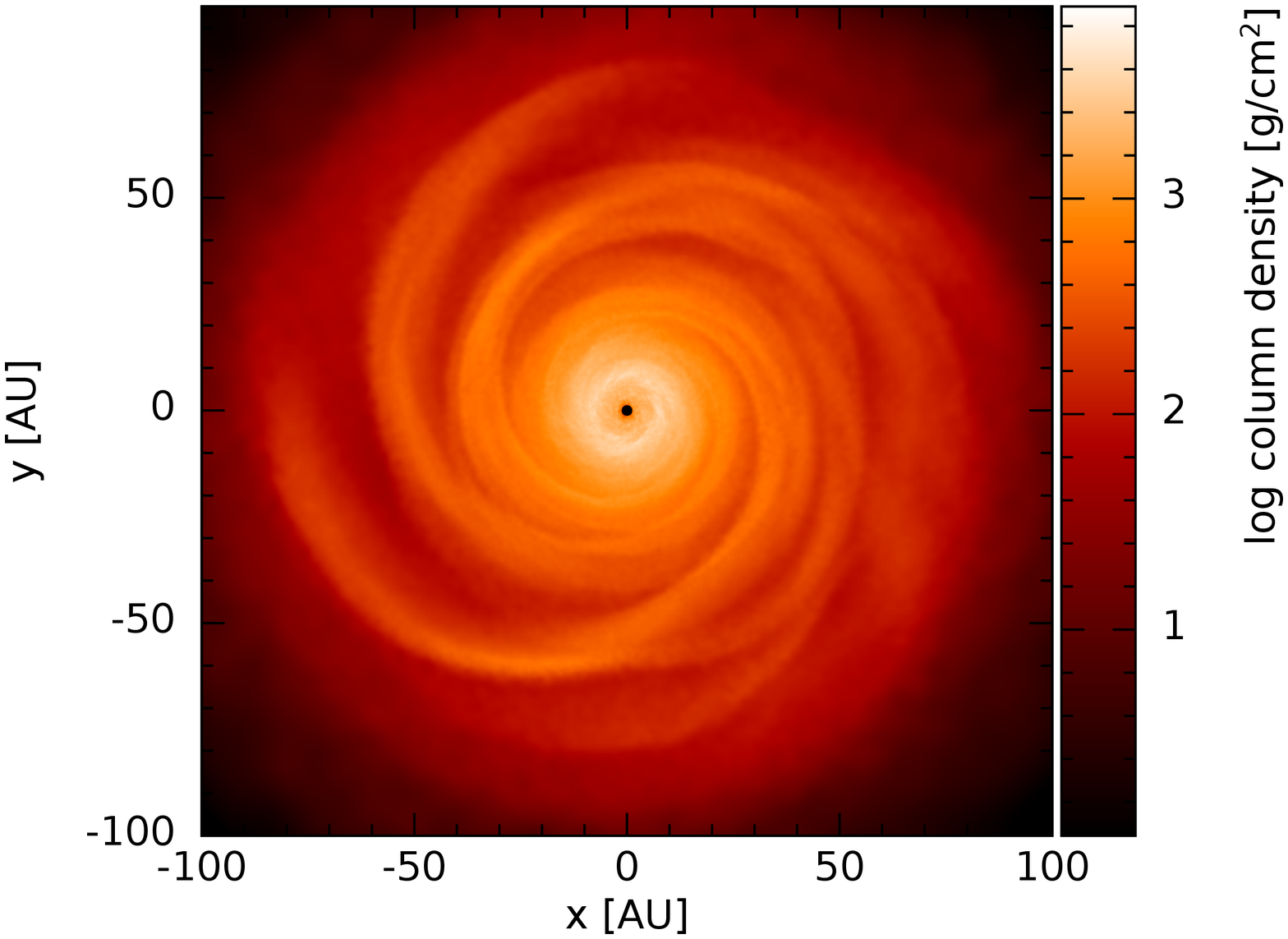}
\includegraphics[width=0.67\columnwidth,angle=-90,origin=br]{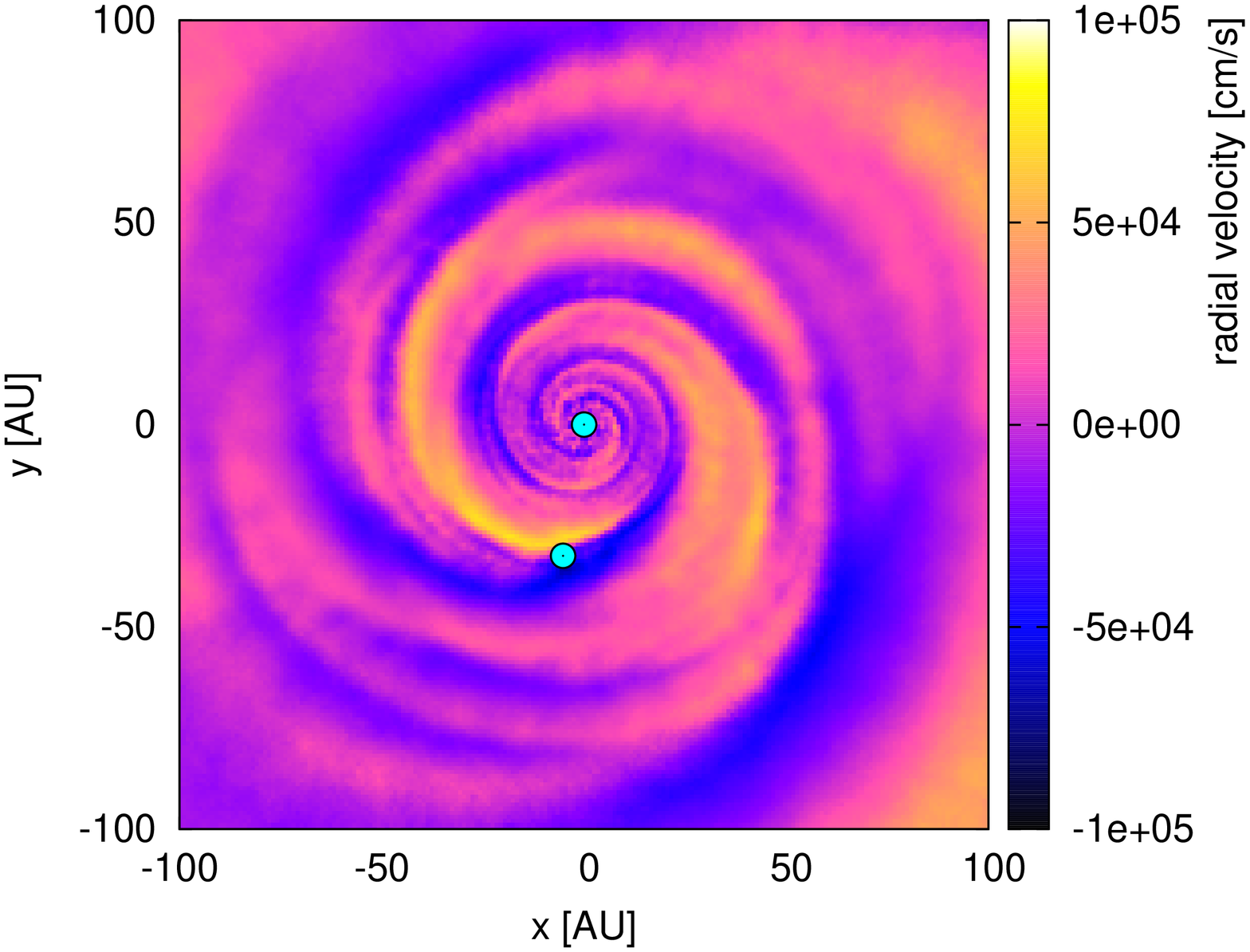}
\includegraphics[width=0.9\columnwidth]{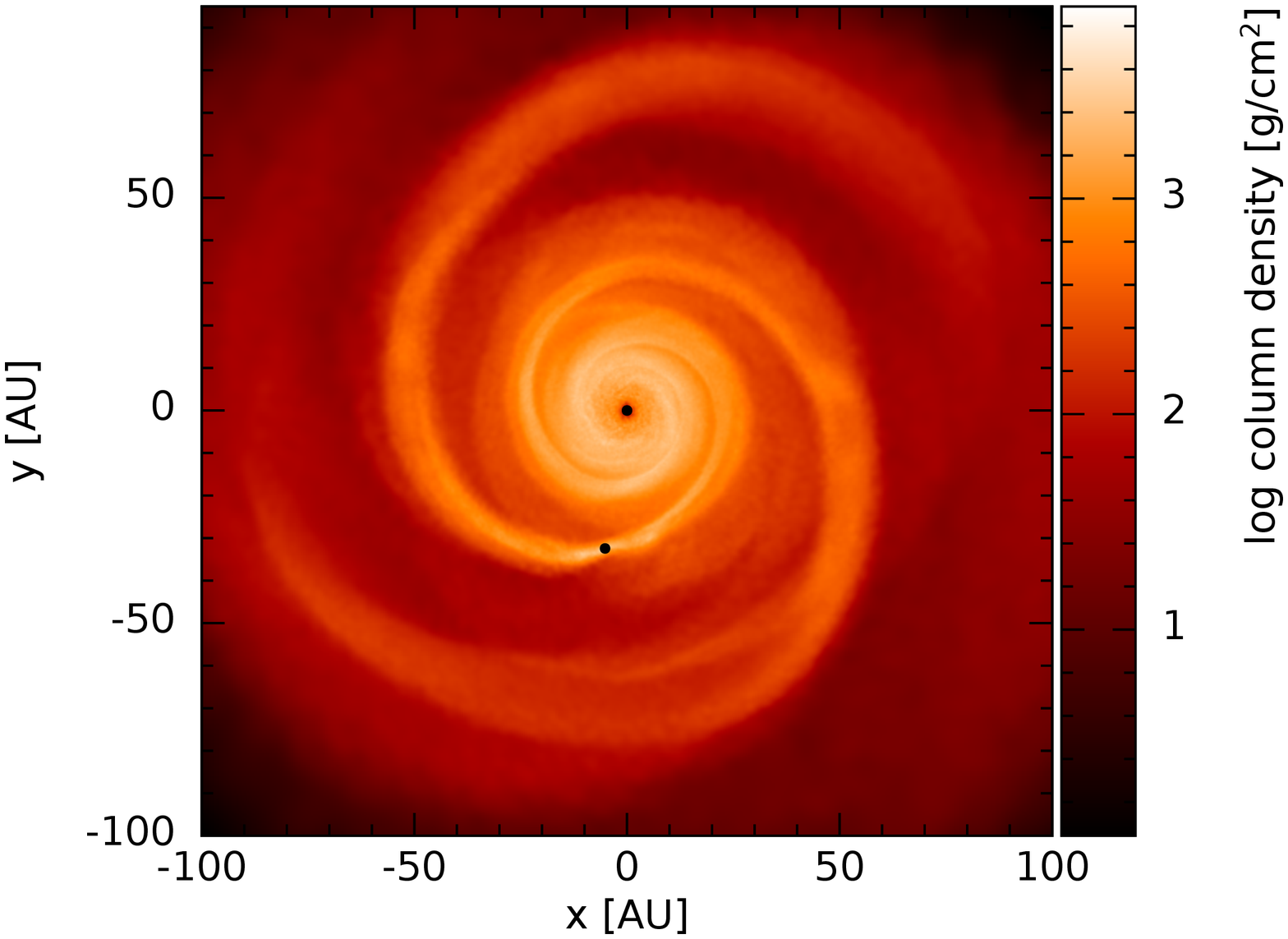}
\includegraphics[width=0.67\columnwidth,angle=-90,origin=br]{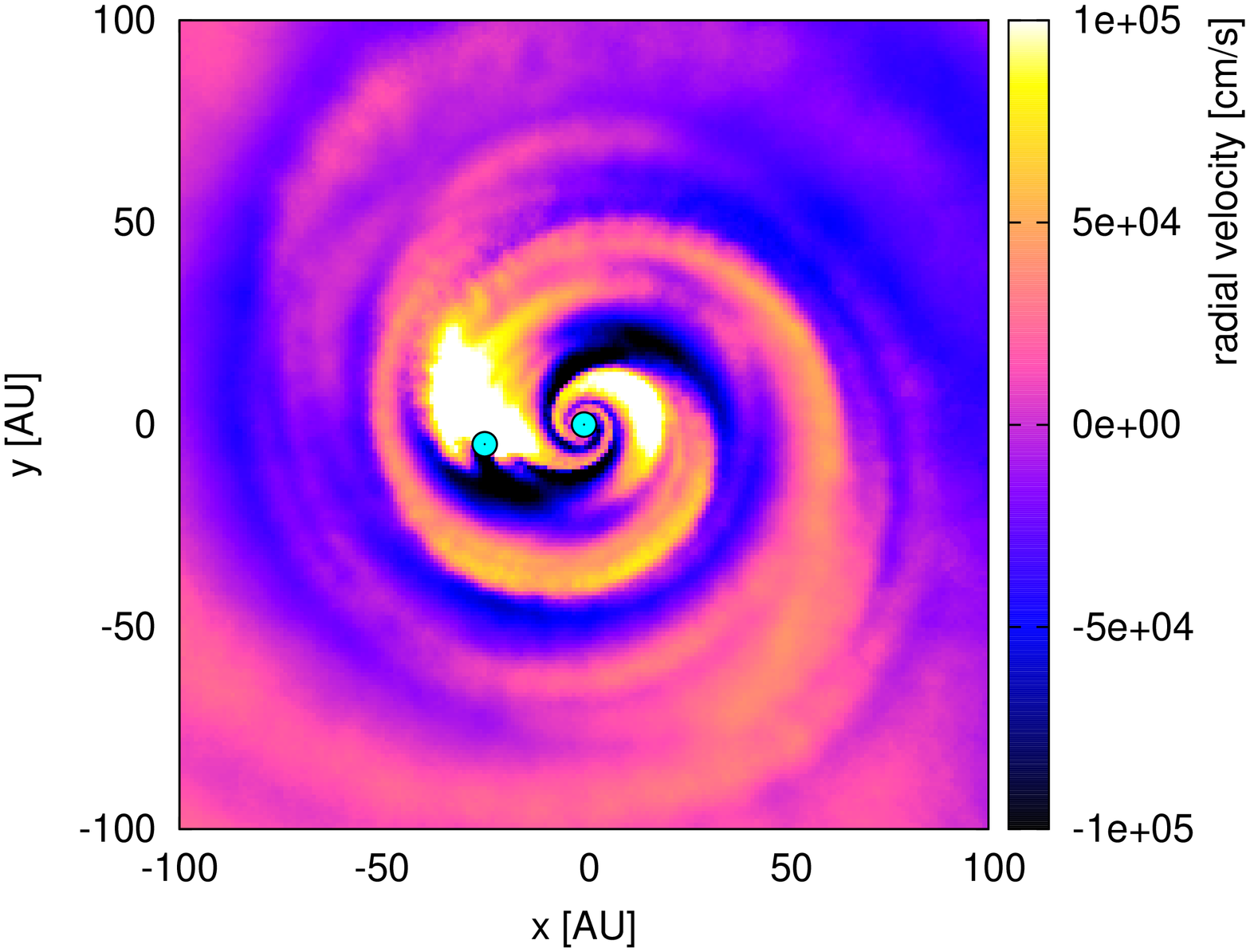}
\includegraphics[width=0.9\columnwidth]{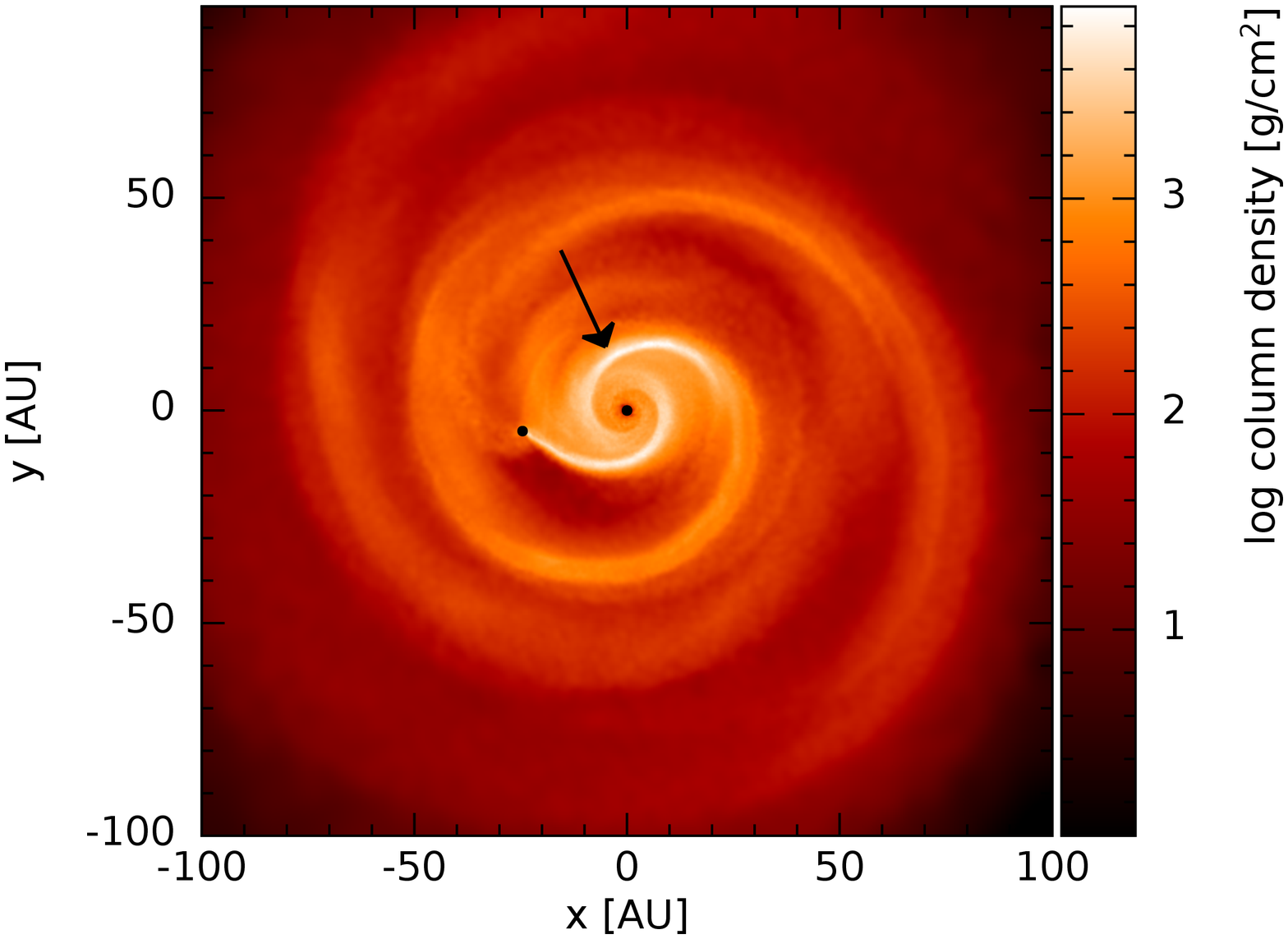}
\includegraphics[width=0.67\columnwidth,angle=-90,origin=br]{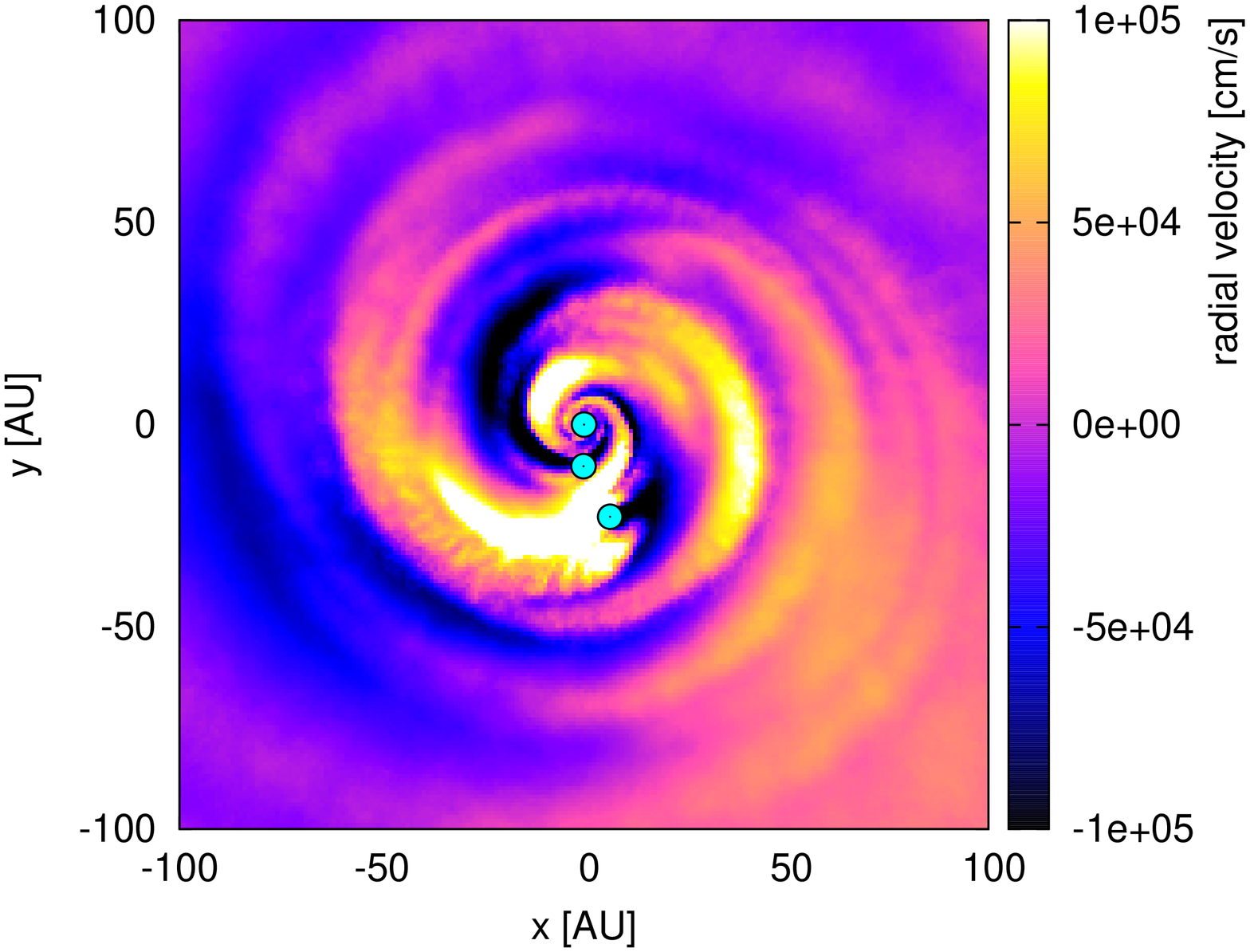}
\includegraphics[width=0.9\columnwidth]{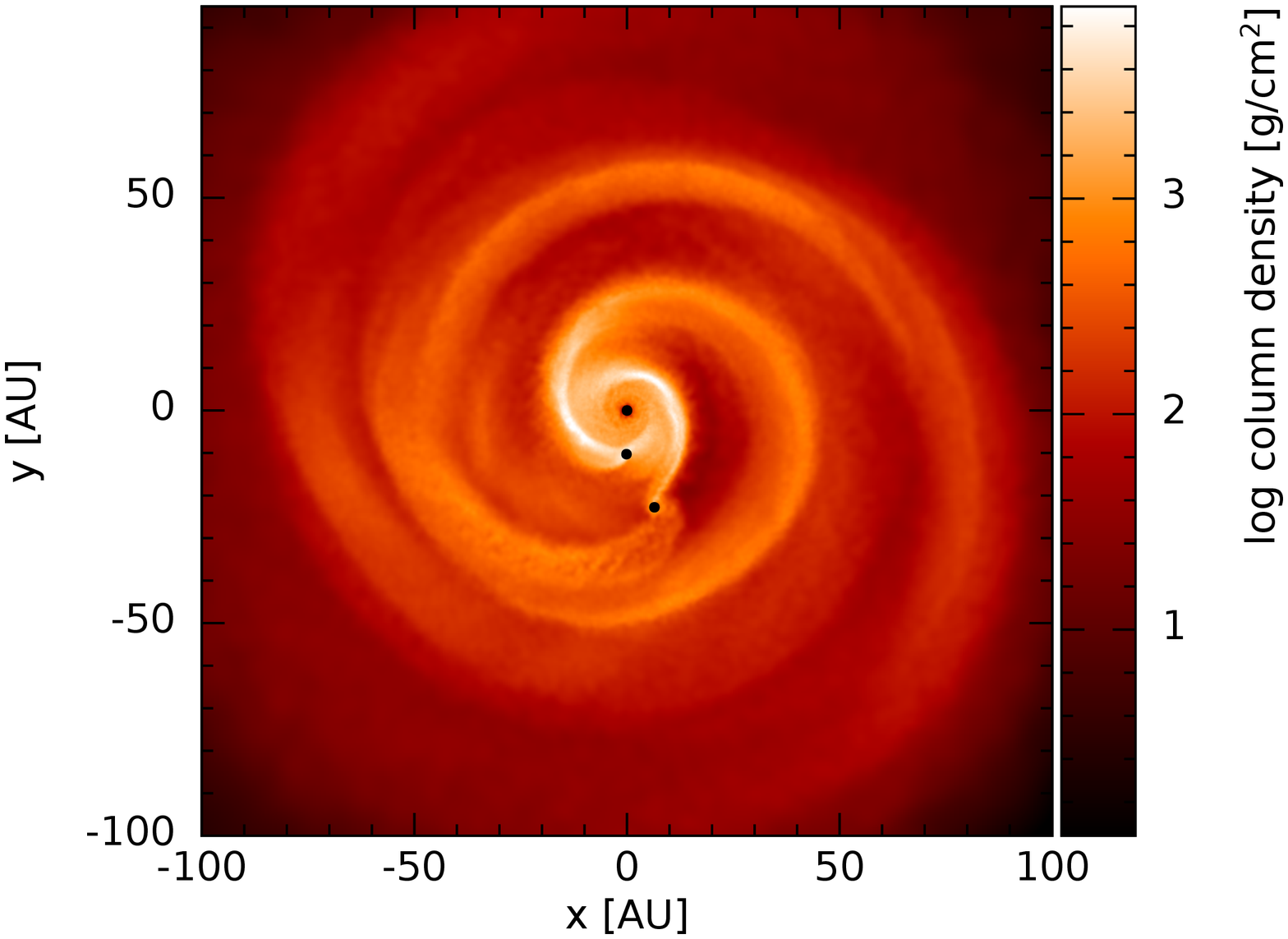}
\vspace{-0.2cm}
 \caption{Radial velocity (left panels) and surface mass density (right panels) rendered images of Simulation 2 before any fragments form (at $t = 0.92$~ORPs; top row), when the first fragment forms (at $t = 1.94$~ORPs; second row), shortly before the second fragment forms (at $t = 2.18$~ORPs; third row) and after the second fragment forms (at $t = 2.20$~ORPs; bottom row).  The disc is initially relatively quiescent.  After the first fragment forms the disc becomes dynamic, causing gas to move more rapidly in both directions.  The inwards movement results in mass piling onto an inner spiral causing it to fragment.  The central star and fragment are given by blue and black dots in the left and right columns, respectively.}
 \label{fig:sim2}
\end{figure*}

\begin{figure*}
\centering \includegraphics[width=0.73\columnwidth,angle=-90]{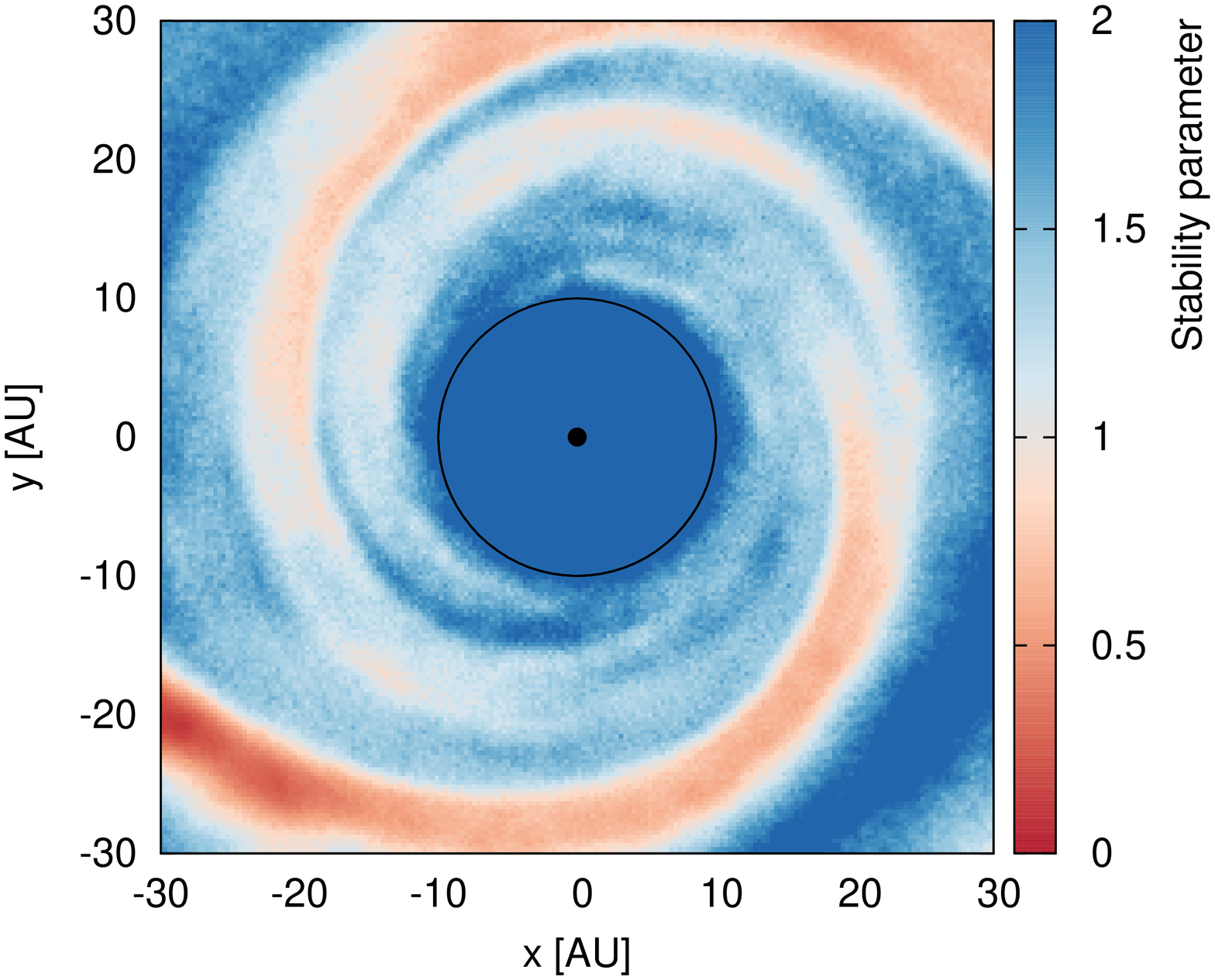}
\includegraphics[width=0.73\columnwidth,angle=-90]{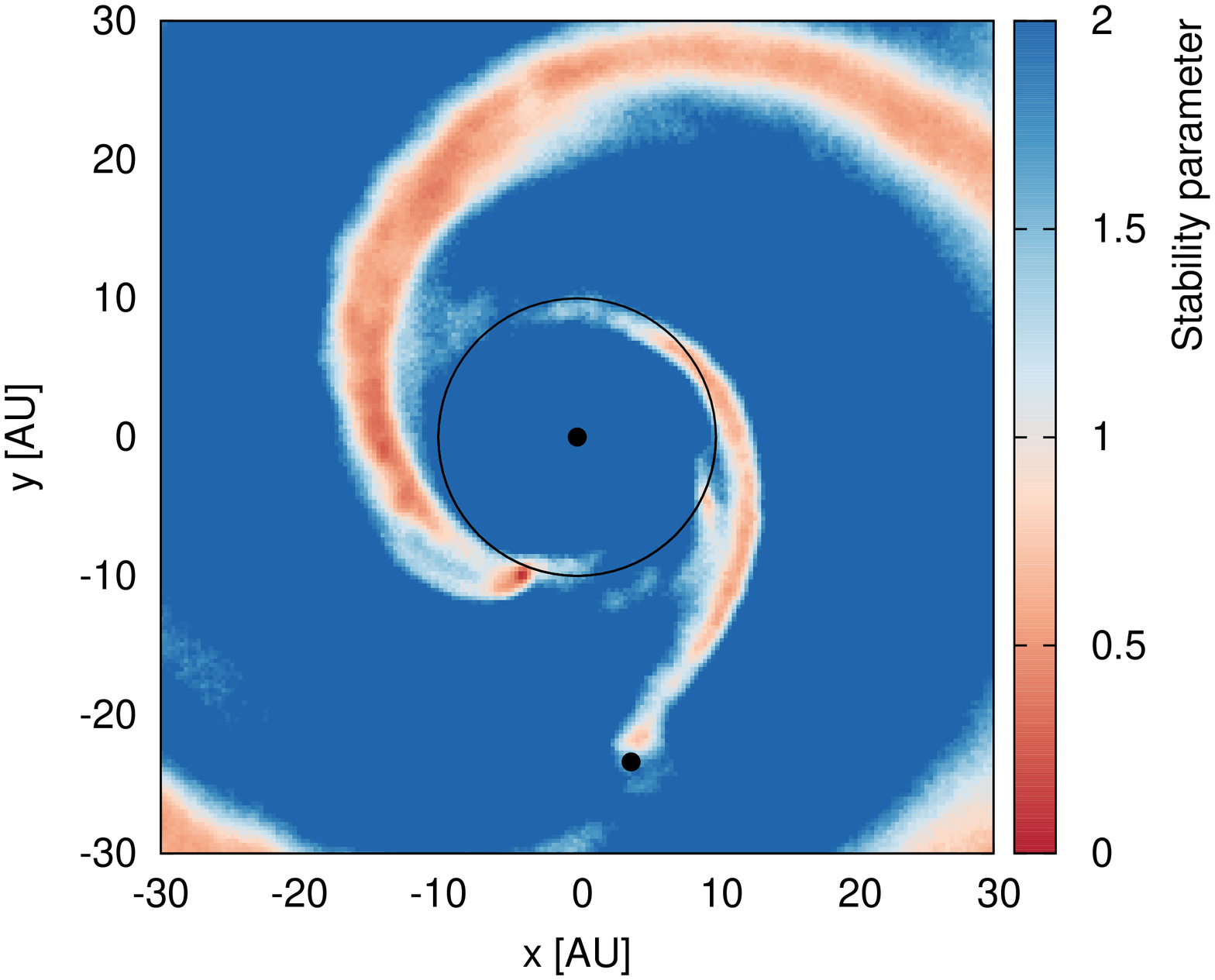}
 \caption{
   Local stability parameter values of the inner disc in Simulation 2 just before the first (left panel) and second (right panel) fragments form (at $t \approx 1.9$ and 2.2 ORPs, respectively).  The black circle indicates the radius at 10 au.  Before the first fragment forms the inner disc is stable and cannot fragment.  However just before the second fragment forms regions of the annulus at $\approx 10$ au are definitely unstable with $Q < 1$, allowing the disc to fragment.  The second fragmentation is unlikely to have occurred without the aid of the mass movement since the stability parameter value is $\gtrsim 1$ at $\approx 10$~au at the earlier time.  The central star and fragment are given by black dots.}
 \label{fig:sim2_Q}
\end{figure*}

\begin{figure*}
\centering
\includegraphics[width=0.73\columnwidth,angle=-90]{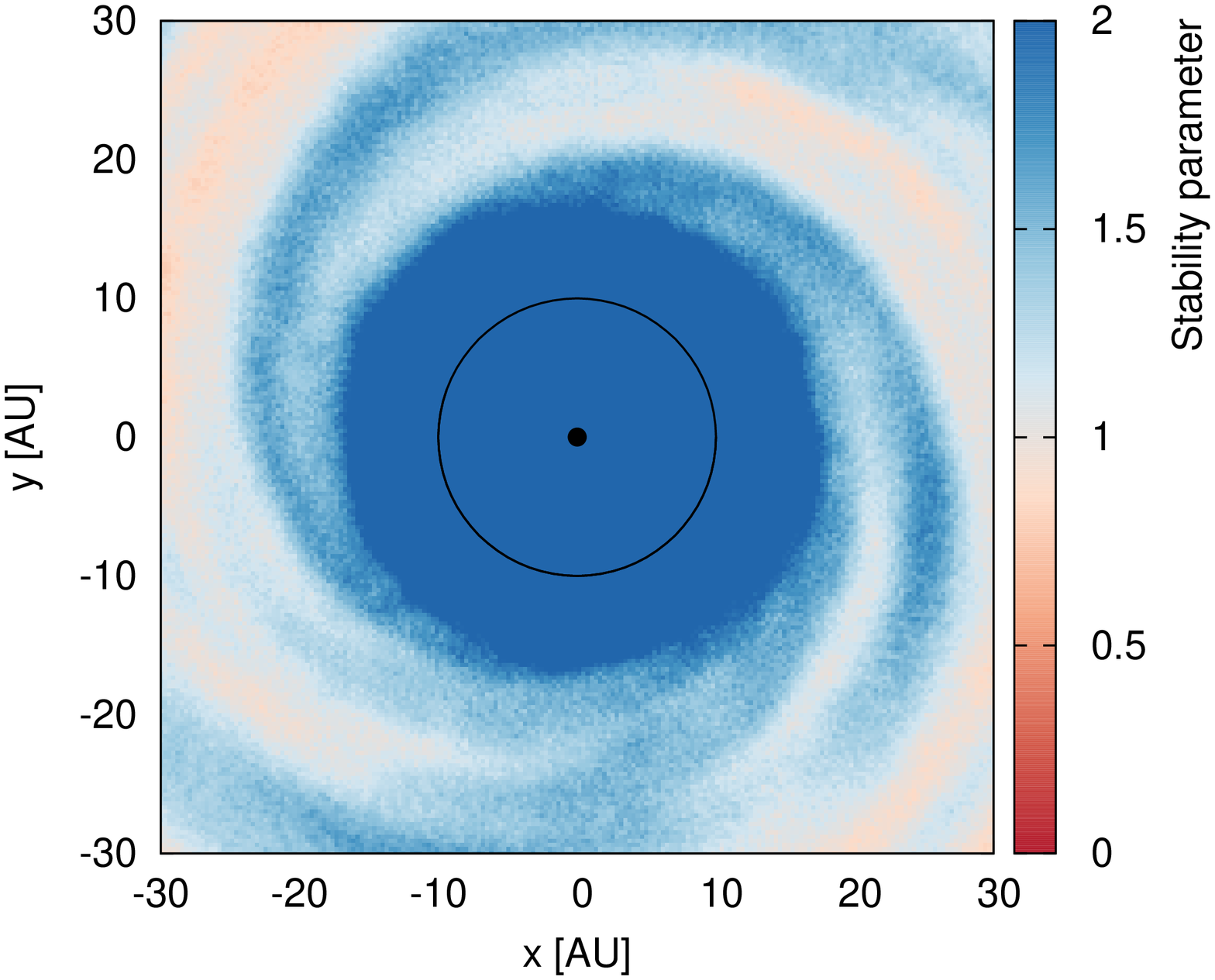}
\includegraphics[width=0.73\columnwidth,angle=-90]{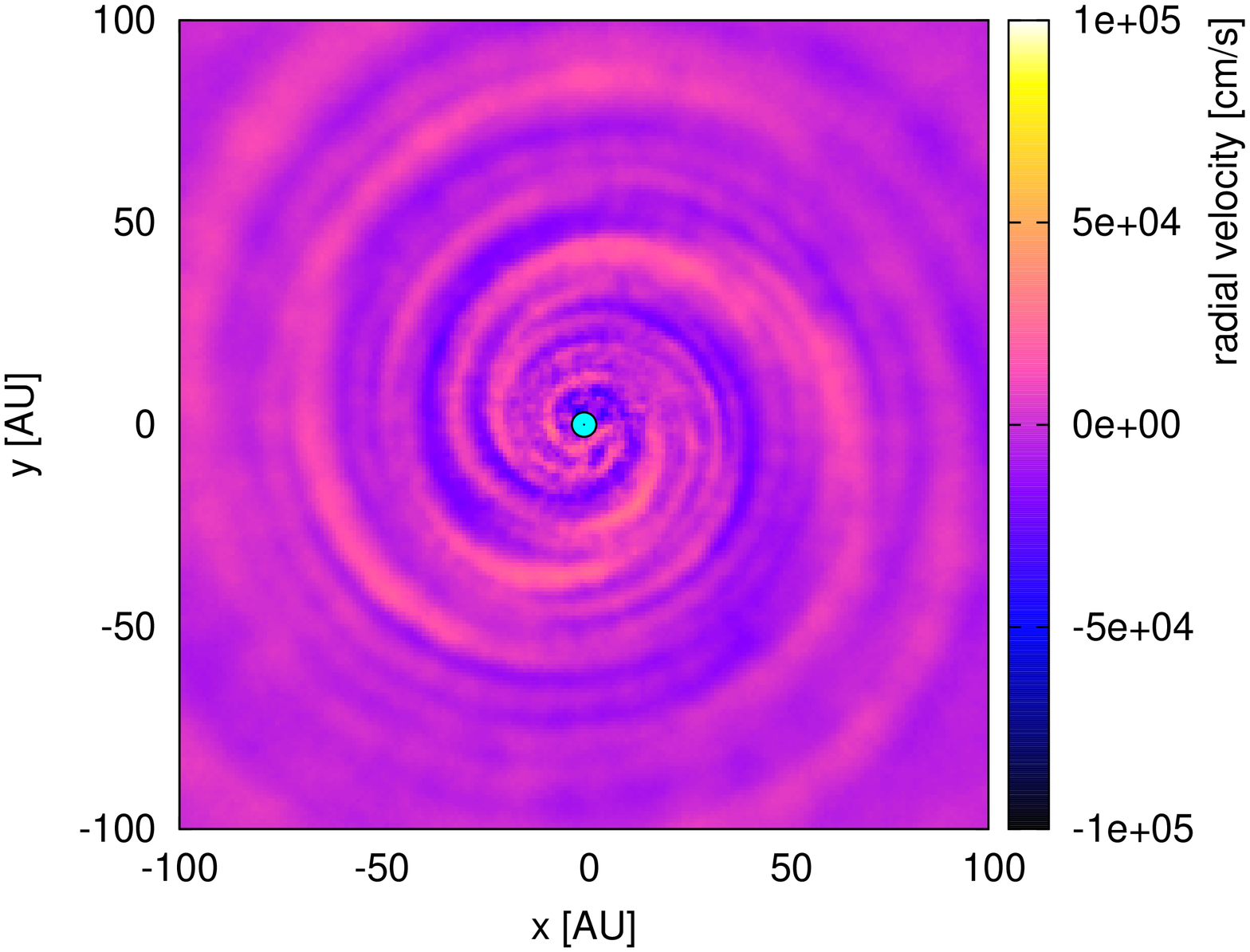}
 \caption{Local stability parameter rendered (left panel) and radial velocity rendered (right panel) plot for the disc at the end of Simulation 2 (at t = 11.1 ORPs) where an annulus of particles has been removed.  The inner disc is stable ($Q > 1$) and does not fragment if the outer fragment doesn't form.  In addition the disc is quiescent without significant mass movement, as it was in the original simulation before any fragments formed (Figure~\ref{fig:sim2}, top left)}
 \label{fig:sim2_kill_Q_vR}
\end{figure*}

Simulation 2 shows a similar behaviour to that of Simulation 1.  The disc becomes unstable but is quiescent before any fragmentation occurs (Figure~\ref{fig:sim2}, top row).  The disc fragments first at $R \approx 33$ au (at $t = 1.94$~ORPs; Figure~\ref{fig:sim2} second row) after which it becomes more dynamic with more mass moving in both directions (Figure~\ref{fig:sim2}, third row).  The inwards movement causes gas to pile onto an inner spiral (indicated by an arrow in Figure~\ref{fig:sim2}) and fragment (at $R \approx 10$ au at $t = 2.2$~ORP; Figure~\ref{fig:sim2} bottom row).  Figure~\ref{fig:sim2_Q} shows that before the first fragment forms the inner disc is stable and is not expected to fragment ($Q > 1$) but at a late time this region is pushed into a state of instability.

As with Simulation 1, to test whether the inner fragment would have formed anyway we restart the simulation at $t = 1.3$~ORPs but remove the particles from the annulus between $32 < R < 34$ au (totalling $0.021 \Msolar$ which is 2.6\% of the SPH particles at that time) so that the formation of the outer fragment is suppressed.  We run the simulations until 11.1 orbits (more than 5 times the time it took until the second fragment formed in the original simulation) and find that no fragments form.  Figure~\ref{fig:sim2_kill_Q_vR} (left panel) shows the stability map at the end of the simulation and shows that the disc is stable, and the radial velocity rendered image of the disc shows it to be calm and quiescent (Figure~\ref{fig:sim2_kill_Q_vR}, right panel), as it was before any fragmentation occurred in the original simulation (Figure~\ref{fig:sim2}, top left).

\section{Discussion}
\label{sec:disc}

Disc fragmentation is known to occur at large radii in young discs.  However the subsequent evolution of a disc immediately after fragmentation and the potential for further fragmentation as a result of a fragment forming is typically understudied when considering this mechanism.  In this paper we show that the movement of mass as a consequence of disc fragmentation at large radii can potentially cause an inner spiral in a disc to become sufficiently dense that it may fragment.  The interaction between the disc gas and the fragment can be thought of as analogous to the theory behind gap-opening by planets.  A region of gas just interior to the fragment transfers angular momentum to the fragment.  Its magnitude is approximately given by \citep[see e.g.][]{Lubow_Ida_migration_review}:

\begin{equation}
|\Delta J|\approx \Big |M a^2 \Omega_f \Big (\frac{M_f}{M_{\star}} \Big)^2 \Big (\frac{a}{x} \Big)^5 \Big|
\label{eqn:ang_mom}
\end{equation}
where $M$ is the mass of the region of gas being considered, $\Omega_f$ and $M_f$ are the angular frequency and fragment mass, respectively, $M_{\star}$ is the mass of the central star, $a$ is the fragment's radial location in the disc and $x$ is the distance between the fragment and the region of gas being considered.  This causes the region of gas to move inwards.  Similarly angular momentum is also transferred from the fragment to the gas just exterior to it causing the gas to move to larger radii.  The inwards motion potentially causes an inner spiral to become sufficiently dense such that it can become unstable and fragment.  This is particularly likely if the inner spiral is at a marginally stable state (with $Q \approx 1$) and just requires an additional push to cause it to become unstable and fragment.  The inwards motion is analagous to a high mass accretion rate through the disc \citep{Hayfield_GI_disc_accretion} or a high infall rate \citep{Vorobyov_Basu_GI_FUOri,Kratter_GI_envelope_accretion,Boley_CA_and_GI,Zhu_GI_migration} which increases the local disc surface density and decreases the local stability, causing fragmentation to occur.

\cite{GI_fragments_ALMA} suggest that mass infall may cause periodic bursts of fragmentation.  It is also possible that the infall causes the first fragment to form, following which further fragmentation is triggered (or accelerated) according to the mechanism in this paper.

It is important to note that once multiple fragments form, the presence of all the fragments, as well as the secular evolution of the disc, will play a part in determining how the disc mass moves.  Indeed, when run for longer, more than two fragments do form.  Previous simulations suggest that a period of chaotic behaviour may occur when multiple objects co-exist in a disc \citep{Pierens_Nelson_multi_planet_mig}.  It is beyond the scope of this paper to characterise the detailed mass movement that may occur in any one particular case since each disc is likely to be dynamically different to one with different disc properties or a slightly varying evolution of fragments.  We stress that the focus here is that the mass movement in a disc is much more dynamic after fragmentation than before, and that this dynamical activity may cause subsequent fragmentation.  We also stress that since the details into the fragmentation of self-gravitating discs have recently been called into question \citep{Meru_Bate_resolution,Meru_Bate_fragmentation,Meru_Bate_convergence,Lodato_Clarke_resolution,Paardekooper_convergence,Rice_cooling_covergence2} with further complications when considering that fragmentation is also somewhat a stochastic process \citep{Paardekooper_shearing_sheet,Hopkins_fragmentation}, the emphasis here is not on fragmentation itself, but the dynamical movement of the disc mass following the initial fragmentation stage, which can then cause an inner spiral to become sufficiently dense such that it can \emph{potentially} fragment.

\cite{Meru_Bate_convergence} showed using their simulations of fragmenting self-gravitating discs using a simplified cooling method that the values of the artificial viscosity parameters used in simulations can result in counterintuitively high dissipation, which may cause fragmentation to be inhibited.  While it remains to be tested exactly how the dissipation affects the fragmentation in discs modelled with radiative transfer, we stress that the inhibition of fragmentation only means that our results are erring on the side of caution rather than over-predicting fragmentation.

In the simulations presented here, since we have modelled a disc more realistically using radiative transfer and by considering the effects of stellar irradiation on the surface temperature of the disc, the initial and subsequent fragmentations take place in the outer and inner regions, respectively.  We point out that if a simulation is carried out using a simplified parameterisation for the cooling (\citealp{Gammie_betacool}; as done so in many previous simulations), the first fragments would form at small radii (since the absolute cooling timescale is shortest there; \citealp{Meru_Bate_fragmentation}) and the subsequent fragmentation would then occur in the outer regions.  While in this case, the subsequent fragmentation would be expected to occur anyway, the fragmentation may occur faster due to it being aided by the movement of mass in the disc (though this time in an outwards direction).  Furthermore, in discs modelled using radiative transfer which may subsequently fragment anyway either interior or exterior to the first fragment irrespective of the gas dynamics, the mass movement may cause the disc to fragment faster by aiding the destabilisation process.

\subsection{In the context of planet formation}
Our results may have important implications for the theories of giant planet formation: while a disc may only initially fragment in the outer regions \citep{Rafikov_SI,Clarke2009_analytical,Boley_CA_and_GI}, planets may well form in the inner parts via gravitational instability \emph{provided} fragmentation has already occurred in the outer disc first.

Core accretion has historically been thought to occur within a disc's lifetime at small radii (up to $\approx 5-10$~AU).  Gravitational instability, on the other hand, is thought to occur at large radii ($\gtrsim 50$~AU; \citealp{Rafikov_SI,Clarke2009_analytical,Meru_Bate_convergence}).  Though a number of factors may influence these radii, e.g. disc mass, stellar mass, metallicity (or equivalently, disc opacity) \citep[e.g.][]{Safronov_CA,Whitworth_Rfrag_Mstar,Cai_etal_RT,Cossins_opacity_beta,Meru_Bate_opacity,Rogers_Wadsley_kappa}, an intermediate region appears to exist ($\approx 10-50$~AU) in which in situ formation by either mechanism is somewhat harder.  Recent advances in growth models at various size scales \citep{Lambrechts_pebble_accretion,Garaud_vel_pdf,Helled_Uranus_Neptune_insitu} have helped to improve the possibility of in situ growth by core accretion at larger distances and the results presented in this paper show that \emph{subsequent planet formation by gravitational instability} may well be a mechanism that may operate to form planets at smaller radii than expected.  Thus both formation models have made progress in the intermediate radial range where no one single formation method has historically been thought to dominate.

\cite{Armitage_Hansen_trigger} first presented the idea that the presence of a planet in a self-gravitating disc may trigger the formation of a fragment in another part of the disc (both interior and exterior to it) where the stability parameter is $\approx 1-2$.  However, these discs were modelled using an isothermal equation of state (which favours fragmentation) and by inserting a planet artificially.  We show that triggered fragmentation occurs even when considering more realistic energetics and when the formation of the first fragment is modelled self-consistently.  Our results are nevertheless consistent with those of \cite{Armitage_Hansen_trigger}.

We also note that the simulations presented here are of high mass discs.  The mass of the fragments when they first form range between $0.7 - 1.2 \MJup$.  However, by the time the second fragments form the first fragments have grown to $\approx 62 \MJup$ in both simulations.  While we expect that mass movement would also take place in lower mass discs, the extent to which the inner spirals become sufficiently dense will be different.  Equation~\ref{eqn:ang_mom} shows that the amount of angular momentum transfer depends not only on the fragment mass but also on the mass of the region of gas being considered (both will ultimately depend on the disc mass).  Therefore, we expect that the mass movement in a lower mass disc will be less violent.  While we have shown the mass movement to be important in high mass discs which are in their very young phase, we present this as a \emph{possible} mechanism by which further fragmentation at smaller radii may occur after an initial fragmentation stage in lower mass self-gravitating discs, where the disc mass is of the order of 10\% of the stellar mass.  However, this needs to be verified with numerical simulations.  Furthermore, in the context of planet formation it is possible that the fragments could lose mass to become planetary mass objects \citep{Boley_tidal_downsizing,Nayakshin_tidal_downsizing}.  However, our simulations use sink particles which can only grow with time so cannot model this in detail.

\section{Conclusions}
\label{sec:conc}

We carry out three-dimensional radiation hydrodynamical simulations of self-gravitating discs to investigate the dynamics of the disc following the initial formation of a fragment.  In particular, we self-consistently explore the formation of the first fragment and the resulting potential for subsequent fragmentation.  We find that the movement of disc mass after the first fragment forms is much more dynamic than prior to fragmentation as the gas radial velocity in parts of the disc can increase by up to a factor of $\approx 10$ in both the inward and outward directions.

The inwards movement of mass can cause an inner spiral arm in a gravitationally unstable disc to become sufficiently dense, pushing it from a stable or marginally stable state into a state of instability such that it may fragment.  This may potentially form fragments at radii smaller than typically expected in protoplanetary discs ($\lesssim 50$~AU).  This subsequent fragmentation may provide a possible method by which planets may form in the intermediate radial region ($\approx 10-50$~AU) where no one in situ formation method has historically dominated.

\section*{Acknowledgments}
We thank Matthew Bate, Cathie Clarke, Donald Lynden-Bell, Lucio Mayer, Eduard Vorobyov and the referee for useful comments.  The calculations reported here were performed using the University of Exeter's SGI Altix ICE 8200 supercomputer and Intel Nehalem (i7) cluster, the university cluster of the computing centre of T\"ubingen and the bwGRiD clusters in Karlsruhe, Stuttgart, and T\"ubingen, and the {\sc brutus} cluster at ETH Z\"urich.  Some disc images were produced using SPLASH \citep{SPLASH}.  FM acknowledges the support through grant KL 650/8-2 (within FOR 759) of the German Research Foundation (DFG).  FM was also supported by the ETH Zurich Postdoctoral Fellowship Programme as well as by the Marie Curie Actions for People COFUND program.  This work has also been supported by the DISCSIM project, grant agreement 341137 funded by the European Research Council under ERC-2013-ADG.

\bibliographystyle{mn2e}
\bibliography{../../allpapers}

\end{document}